\documentclass[aps,prr,longbibliography,floatfix,twocolumn,showpacs,amsmath,amssymb,superscriptaddress,titlepage,natbib]{revtex4-1}
\usepackage{graphicx} 
\usepackage{subfigure}
\usepackage{multirow}
\usepackage{fancyhdr}
\usepackage{longtable}
\usepackage{parskip}
\usepackage[T1]{fontenc}
\usepackage{dcolumn} 
\usepackage{bm}       
\usepackage{amsfonts}  
\usepackage{amsmath}   
\usepackage{amssymb}   
\usepackage[]{hyperref}

\usepackage[utf8]{inputenc}
\usepackage[T1]{fontenc}
\usepackage{mathptmx}
\usepackage{etoolbox}
\usepackage{xcolor}

\begin{document}
	
\title{Synchronization of  the time-delayed Kuramoto model in a regular network}

\author{Sara Ameli}%\email{es.mahdavi@iasbs.ac.ir}
\affiliation{Forschungszentrum Julich GmbH, Peter Grunberg Institut (PGI-14): Neuromorphic Compute Nodes, Julich, 52425, Germany,}
%\affiliation{Faculty of Electrical Engineering and Information Technology, RWTH Aachen University, Aachen, 52056, Germany.}

\author{Esmaeil Mahdavi}%\email{es.mahdavi@iasbs.ac.ir}
\affiliation{ Department of Physics, Institute for Advanced Studies in Basic Sciences (IASBS), Zanjan 45137-66731, Iran}

\author{Mina Zarei}%\email{mina.zarei@iasbs.ac.ir}
\affiliation{ Department of Physics, Institute for Advanced Studies in Basic Sciences (IASBS), Zanjan 45137-66731, Iran}

\author{Farhad Shahbazi} %\email{shahbazi@cc.iut.ac.ir }
\affiliation{ Department of Physics, Isfahan University of Technology, Isfahan 84156-83111, Iran}

\begin{abstract}
This study investigates the impact of delayed coupling on the global and local synchronization of identical coupled oscillators residing in a ring. Utilizing the Kuramoto model, we examine the effects of delayed coupling on collective dynamics. Our analytical and numerical results reveal distinct synchronization behaviors across various time delay regimes, including fully synchronized states, helical patterns, dynamically incoherent states, and random phase-locked states. We identify the regime in which time delay inhibits synchrony, which aligns with theoretical predictions derived from stability analysis. In the region between the synchrony-possible and synchrony-forbidden zones, a coexistence of synchronized and unsynchronized dynamics is observed, referred to as Chimera states. 
We ascertain that the type of Chimera states present is moving-turbulent.
\end{abstract}

\maketitle

\section{Introduction}
\label{sec1}
Synchronization is the coordinated activity of numerous agents that allows systems to function cohesively across various fields, including biology, technology, and social systems~\cite{strogatz2012sync,nijmeijer2003synchronization,mirollo1990synchronization,tyson1973some,blasius1999complex,pikovsky2003synchronization}. Understanding synchronization is important because it demonstrates how individual components work together to create collective behavior. Significant research has focused on developing mathematical and computational models to tackle the complexities of synchronized behavior. These models have taken into account various factors, such as coupling structures, frequency distributions, noise, inertia, and time delays~\cite{arenas2008synchronization,pecora1998master,nakao2007noise, schuster1989mutual,mahdavi2025synchronization, ameli2024two, ameli2022low}. 
Incorporating these elements not only enhances theoretical understanding but also aids in the development of effective control strategies, optimizes performance in engineered systems, and provides insights into collective behaviors in both natural and artificial environments.

The Kuramoto model serves as a foundational framework for exploring phase synchronization phenomena, utilizing a straightforward mathematical approach to analyze the dynamics of coupled oscillators. Originally designed for fully connected networks, the model has been extended to encompass more complex topologies, offering valuable insights into the principles that govern synchronization dynamics. Among these configurations, regular networks stand out due to their nodes having an equal number of connections. This structured connectivity is particularly significant because it fosters localized interactions among oscillators, thereby improving our understanding of how local dynamics can influence overall synchronization.

Regular networks serve as a strong basis for understanding the fundamental mechanisms of synchronization. They allow researchers to investigate how uniform connectivity influences collective behavior in oscillatory systems. This framework has practical applications; for example, certain arrays of Josephson junctions can be effectively modeled as a Kuramoto-type system with nearest-neighbor coupling~\cite{daniels2003synchronization,watanabe1996dynamics}. Existing studies have predominantly focused on ring networks, a specific type of regular network in which each node connects symmetrically to some neighbors. While this one-dimensional topology may appear to be simple, it reveals a range of intriguing dynamical features. The system's symmetry enables various dynamically stable stationary states, each representing distinct forms of collective behavior. Studies have indicated that ring configurations can display multiple synchronized states, each with unique winding numbers. The winding number, defined as the number of times the oscillator's phases complete a full cycle encircling across the ring, serves as a key indicator of phase-locking. Importantly, these states differ not only in stability but also in the size of their basins of attraction, with higher winding numbers generally associated with smaller basins of attraction. Additionally, some unstable states have also been identified~\cite{wiley2006size, article, tilles2011multistable, denes2019pattern,denes2019predictability}.

Incorporating time delays into the Kuramoto model is essential for accurately representing real-world interactions, where communication between oscillators involves finite transmission times. Recognizing these delays greatly enhances our understanding of synchronization in various applications, such as neural networks, power grids, arrays of lasers, electronic circuits, and microwave oscillators, where interactions are inherently non-instantaneous
~\cite{kerszberg1990synchronization,waibel2013phoneme,kozyreff2000global,reddy2000experimental,yeung1999time,taher2019enhancing}. This perspective allows for a more effective exploration of synchronization while addressing its complexities. Time-delayed systems are mathematically equivalent to infinite-dimensional systems because the inclusion of delay adds an extra dimension to the phase space. As a result, the current state of the system can be influenced by its historical state. 
This added complexity ultimately limits the predictability of the system's dynamics~\cite{wernecke2019chaos}. 
The introduction of time delays leads to various synchronization phenomena, such as multistability, asynchronous behaviors, and the emergence of chimera states~\cite{schuster1989mutual,yeung1999time,choi2000synchronization, ameli2021time}. 
In these chimera states, certain oscillators synchronize their behavior while others do not, highlighting the complex dynamics that arise from delayed interactions. This intricate interplay emphasizes the importance of time delays in influencing collective behavior within coupled oscillator systems. Consequently, this paper focuses on the synchronization of Kuramoto oscillators with time delays in regular networks.
The impact of time delay on synchronization in ring networks has been investigated through both theoretical and experimental studies~\cite{earl2003synchronization,denes2021synchronization,matias1997observation}. The stability condition for phase-locked states in ring networks of identical phase oscillators with delayed coupling has been established, and the attraction basins' sizes have been analyzed~\cite{earl2003synchronization,denes2021synchronization}. 

 In our previous work, we investigated the synchronization of small-world networks composed of identical coupled phase oscillators using Kuramoto interactions and uniform time delays. For specific intrinsic frequencies and coupling constants, we observed an enhancement of synchronization across a range of time delays, as well as a discontinuous transition from a partially synchronized state with defect patterns to a glassy phase. Additionally, we identified chimera states~\cite{ameli2021time}. In this paper, we derive stability conditions for synchronized states in regular rings with $k$ neighbors using linear stability analysis. We analyze both analytically and numerically the frequency phase-locked solutions for fully synchronized states. We notice that the time delay affects the winding number. Additionally, we observe the emergence of turbulent chimera states. 

The paper is organized as follows: Section~\ref{model} offers an introduction to the mathematical tools to investigate the phase-locked solutions in the system and evaluate the stability of fully synchronized states based on the phase-locked frequency. Section~\ref{results} presents our numerical findings, highlighting the effects of time delay on the winding number. We then explore chimera states, analyzing their properties and discussing the implications of these different solutions. Finally, we summarize our results in Section~\ref{conclusion}.

\section{Model and method}
 \label{model}

We study the phase dynamics of a network including $N$ identical coupled oscillators with intrinsic frequencies of $\omega_0$, and uniform time delay interaction is described by the delayed Kuramoto model:
\begin{equation}
\frac{d\theta'_i(t')}{dt'}=\omega_0+\frac{K}{k_i}\sum_{j=1}^N a_{ij}\sin\big(\theta'_j(t'-\tau')-\theta'_i(t')\big),~~~~~~~i = 1,..N. 
\label{main}
\end{equation}
where $k_i$, $a_{ij}$, and $\tau'$ denote the number of nearest neighbors of the $i$'th oscillator, the adjacency matrix elements, and time delay, respectively. Typically, the intrinsic angular velocity of oscillators can be selected from a distribution; however, our focus here is exclusively on investigating the impact of time delay. Therefore, the oscillators are considered completely identical for this study.
In this work, we examine a circular regular network topology (ring), where all nodes have the same number of connections from left and right. Therefore,  $k_i$ is the same for all the nodes i.e. $k_i = k$.

Defining $\theta_i(t')= \theta'_i(t')-\omega_0 t'$ (rotating frame) and  the dimensionless time $t = Kt'$, Eq. \eqref{main} transforms to the following form:
\begin{equation}
\frac{d\theta_i(t)}{dt}=\frac{1}{k_i}\sum_{j=1}^N a_{ij}\sin\big(\theta_j(t-\tau)-\theta_i(t)-\frac{\omega_0}{K}\tau\big),
\label{mainEq2}
\end{equation}
where $\tau=K\tau'$ is the dimensionless time delay. From now on, we set $K = 1$, and investigate the impact of the phase shift $\omega_0\tau$ on the dynamics.

To quantify the degree of synchrony in the network, we employ the order parameter $r(t)$, defined as:
\begin{equation}
r(t)=\frac{1}{N}|\sum_{j=1}^Ne^{i\theta_j(t)}|,
\end{equation}
where $r=1$ represents complete synchronization, while $r=0$ indicates an incoherent dynamical state, randomly frozen state, or a regular phase-locked pattern. 
The long-time order parameter $(r_{\infty})$ is measured and averaged over a time
window {$\Delta t$} after the system reaches its stationary state:  
\begin{equation}
r_{\infty}=\lim_{\Delta t\rightarrow\infty}\frac{1}{\Delta t}\int_{t_s}^{\Delta t+t_s}r {dt},
\label{rinf}
\end{equation}
here, $t_s$  is the time after which the system's dynamics have reached a stationary state.

The average collective angular velocity in the rotating frame can be calculated as:

\begin{equation}
\Omega=\frac{1}{N}\sum_{i=1}^N\lim_{\Delta t\rightarrow\infty}\frac{1}{\Delta t}\int_{t_s}^{\Delta t+t_s}\omega_i {dt}.
\label{Omega}
\end{equation}
 In our study, we set \textbf{$t_s = 5\times 10^5$} and the averaging time window $\Delta t = 3\times 10^4$. 
 
To assess the coherence between oscillators $i$ and $j$, we employ the correlation matrix to determine the local phase configuration. The correlation matrix element $D_{ij}$ is defined as:
\begin{equation}
D_{ij}=\lim_{\Delta t\rightarrow\infty}\frac{1}{\Delta t}\int_{t_s}^{\Delta t +t_s}\cos\big(\theta_i(t)-\theta_j(t)\big){dt}.
\label{D}
\end{equation}
Here $-1\leq D_{ij}\leq 1$, where $D_{ij}=1$, $D_{ij}=-1$ denote in-phase and  anti-phase states, respectively. In the case of no coherence between the two oscillators, one finds $D_{ij}=0$.

The local order parameter $R_{i}(t)$ defined as 
\begin{equation}
R_{i}(t)=\frac{1}{2m+1}|\sum_{j=i-m}^{i+m}e^{i\theta_j(t)}|,
\label{r-local}
\end{equation}
indicates the level of synchronization of each node with its $2m$ nearest neighbors.

\section{Results and discussion}
\label{results}
To analyze the synchronization behavior, we employ both numerical and analytical techniques. Our primary focus is investigating how the interplay between network structure and time delay influences global and local synchronization properties. This paper considers a regular network with $N=1000$ and degree $k=10$. We take  $\omega_{0}=\frac{\pi}{2}$ for the intrinsic frequencies of the phase oscillators in the Kuramoto model.

\subsection{Linear stability analysis}

We conduct an analysis of the stability conditions pertinent to fully synchronized states through the utilization of linear stability analysis. By introducing minor perturbations around these synchronized states, we ascertain the stability boundaries corresponding to various time delays. 
For phase-locked solutions in the fully synchronized state, the phase of the oscillators after reaching the stationary state ($\theta^{*}(t)$) can be written as 
\begin{equation}
{\theta}_{i}^{*}(t)=\Omega_{f} t+\beta, 
\label{phasestar}
\end{equation} 
where $\Omega_{f}$ and $\beta$ are the frequency of the fully synchronized state and a constant, respectively.  Inserting Eq.~ \eqref{phasestar} in Eq.~ \eqref{mainEq2}, gives the frequency $\Omega_{f}$ which is the roots of the function $F({\Omega_f)}$

\begin{equation}
F({\Omega_f})=\Omega_{f} + \sin({\Omega_f}\tau+\frac{\omega_{0}\tau}{K}).
\label{fomega}
\end{equation} 
Since the network is regular, $k_i =k=10 $ is uniform across all oscillators. To check the stability of the obtained solutions, we introduce small perturbations to the phase of the oscillators in the stationary state:

\begin{equation}
{\theta_i}(t)={\theta}_{i}^{*}(t)+\xi_{i}(t)=\Omega_{f} t+\beta+\xi_{i}(t). 
\label{dist}
\end{equation} 

Inserting the ansatz \eqref{dist} to Eq. \eqref{mainEq2}, and linearizing this equation one finds

\begin{equation}
\dot{\boldsymbol{\xi}}=\cos({\Omega_f}\tau+\frac{\omega_{0}\tau}{K}) (\boldsymbol{\xi}^{\tau}-\boldsymbol{\xi})-\frac{1}{ k } \cos({\Omega_f}\tau+\frac{\omega_{0}\tau}{K}) \boldsymbol{L}\boldsymbol{\xi}^{\tau},
\label{xi}
\end{equation} 
where $\dot{\boldsymbol{\xi}}=\begin{pmatrix}
    \dot{\xi}_{1}(t)\\
    \vdots\\
    \dot{\xi}_{N}(t)
\end{pmatrix}$, $\boldsymbol{\xi}=\begin{pmatrix}
    \xi_{1}(t)\\
    \vdots\\
    \xi_{N}(t)
\end{pmatrix}$, $\boldsymbol{\xi}^{\tau}=\begin{pmatrix}
    \xi_{1}(t-\tau)\\
    \vdots\\
    \xi_{N}(t-\tau)
\end{pmatrix}$, 
and $\boldsymbol{L}$ is the Laplacian matrix ($l_{ij}=\delta_{ij}k_{i}-a_{ij}$).
Diagonalizing the Laplacian matrix, Eq.~\eqref{xi} can be rewritten as 

\begin{equation}
\dot{\boldsymbol{\eta}}=\cos({\Omega_f}\tau+\frac{\omega_{0}\tau}{K}) (\boldsymbol{\eta}^{\tau}-\boldsymbol{\eta})-\frac{1}{k } \cos({\Omega_f}\tau+\frac{\omega_{0}\tau}{K}) \boldsymbol{\Lambda}\boldsymbol{\eta}^{\tau}.
\label{xi2}
\end{equation}
Here, $\boldsymbol{\Lambda}$ is the diagonalized Laplacian matrix. It can be shown that the solutions of Eq.~\eqref{xi2} are in the form 
${\eta_{i}}\sim \exp{(z_{i}t)}$, where the exponents $z_{i}$ are the solutions of  the following equation
\begin{equation}
z_{i}-\cos({\Omega_f}\tau+\frac{\omega_{0}\tau}{K}) (e^{-z_{i}\tau}-1)-\frac{1}{ k } \cos({\Omega_f}\tau+\frac{\omega_{0}\tau}{K}) {\lambda_{i}}e^{-z_{i}\tau}=0,
\label{zi}
\end{equation}
in which $\lambda_i$'s are the eigenvalues of the Laplacian matrix $L$. The condition for the stability of the solution is that the exponents $z_{i}<0$ for all $i=1,..., N$, otherwise the solution is unstable.

\begin{figure}
\includegraphics[width=\columnwidth]{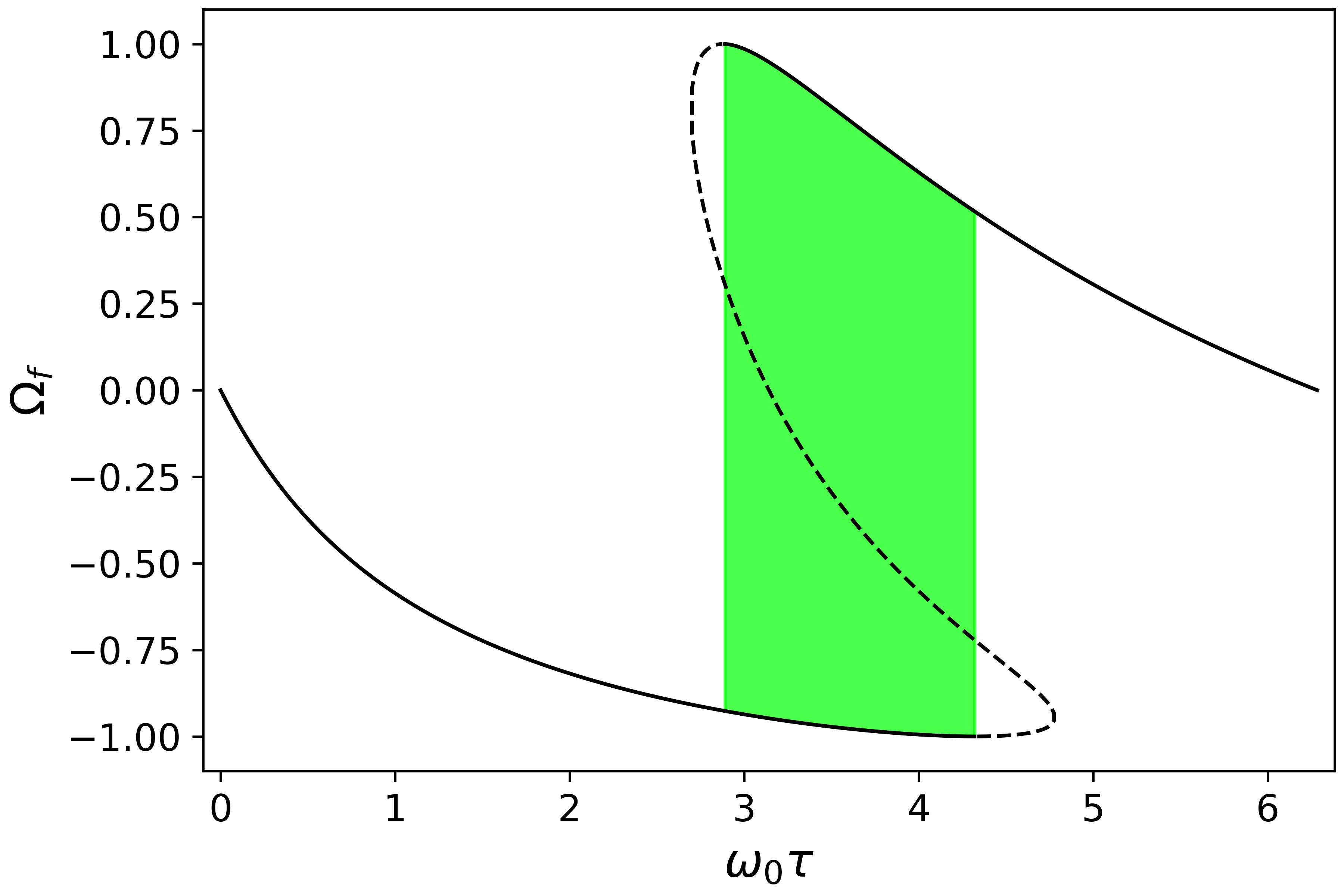}
\caption{(Color online) Angular velocity of the fully synchronized state versus {$\omega_0\tau$} for a regular network of phase oscillators with degree $k=10$ and $\omega_0=\pi/2$. Solid and dashed lines represent the stable and unstable solutions, respectively.  The green area shows the bi-stability region.} 
\label{fig1:analytic}
\end{figure}

Fig.\ref{fig1:analytic} represents the frequency of the fully synchronized solutions obtained by the above procedure for different time delay values.  In this figure, the unstable solutions have been shown as a dashed line. There are two stable branches for the fully synchronized solutions in the range of shown time delays. These two branches coexist in the interval $2.89 \lesssim \omega_{0}\tau\lesssim 4.32$, indicating 
bi-stability in this interval.

%%%%%%%%%%%%%%%%%%%%%%%%%%%%%%%%%Fig1
%%%%%%%%%%%%%%%%%%%%%%%%%%%%%%%%%Fig2
\begin{figure*}
\centering
\subfigure[]{\includegraphics[width=\columnwidth]{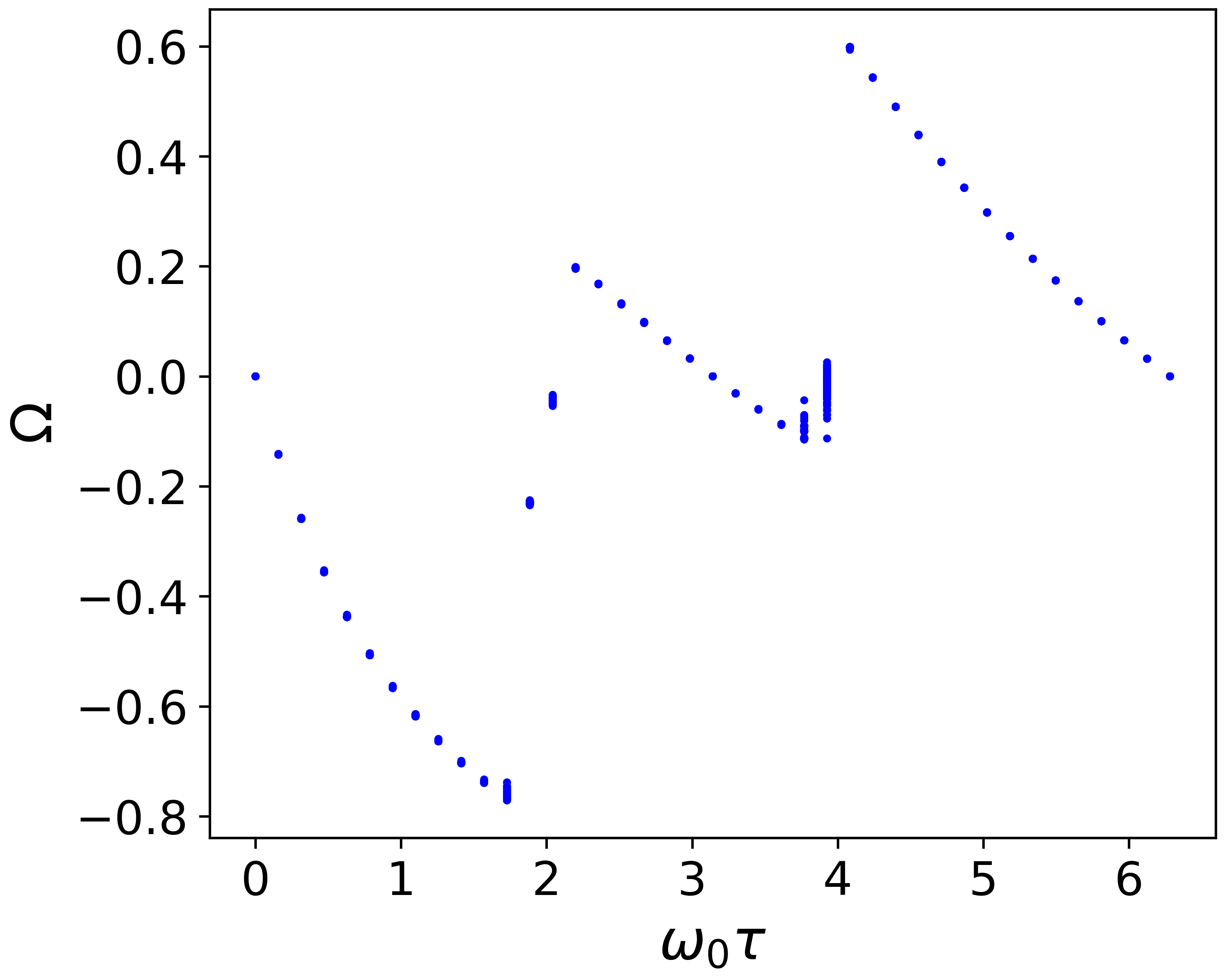}}
\subfigure[]{\includegraphics[width=\columnwidth]{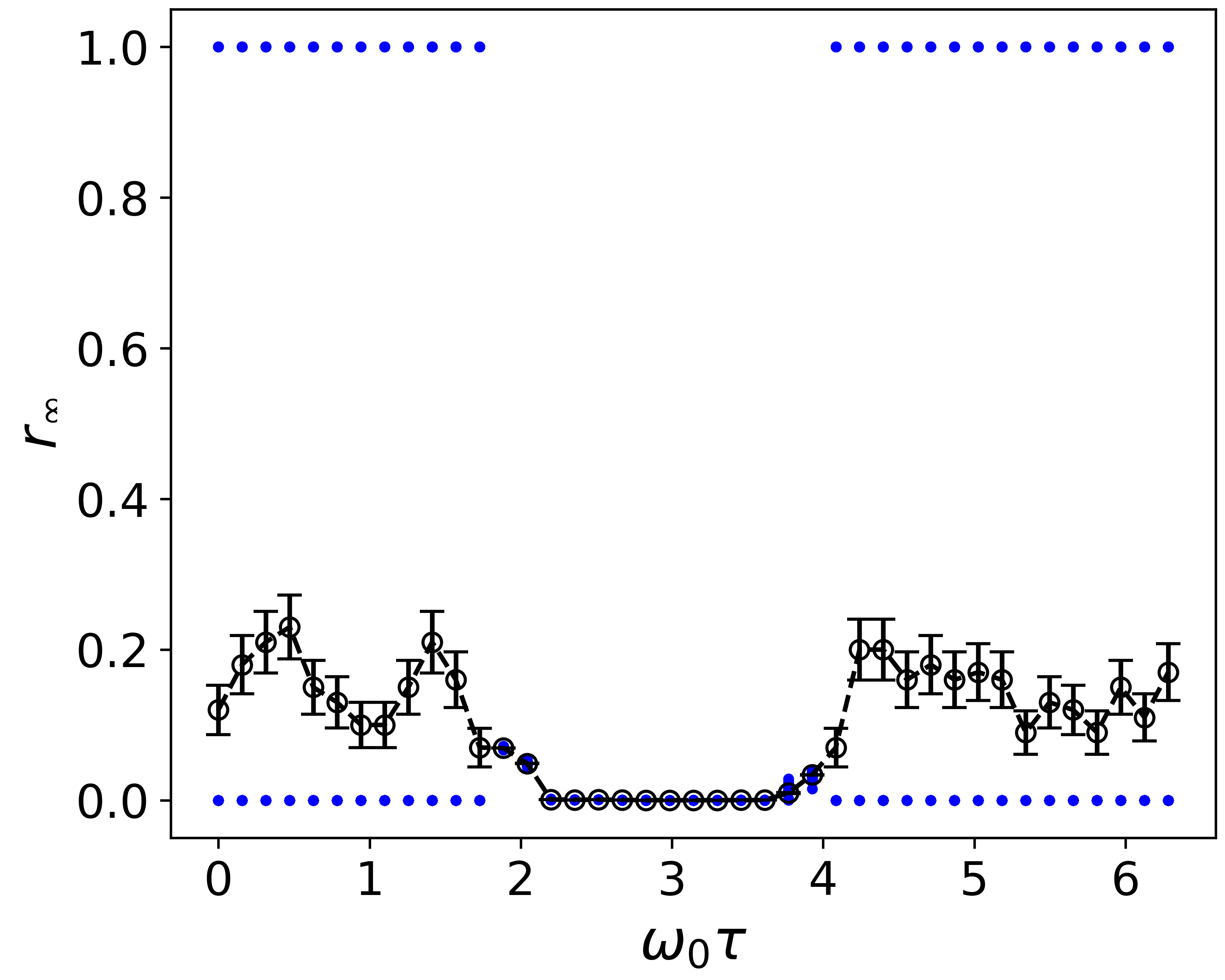}}
\caption{(Color online) (a):  Long-time averaged angular velocity of oscillators $\Omega$, and (b) long-time averaged order parameter ($r_{\infty}$) versus $\omega_0\tau$. The size of the network is $N=1000$, its degree is $k=10$ and the intrinsic angular velocity of the oscillators is $\omega_0 = \pi/2$. In both panels, the blue dots denote the results of realizations with $100$ different random initial phase distributions. The open circles in panel (b) denote the average of  $r_{\infty}$ over the $100$ realizations and the error bars show the standard error of means. 
 }
\label{fig2:rw}
\end{figure*}
%%%%%%%%%%%%%%%%%%%%%%%%%%%%%%%%%%%

%%%%%%%%%%%%%%%%%%%%%%%%%%%%%%%%%%%%%%%%Fig3
\begin{figure*}[]
\centering
\includegraphics[width=\textwidth]{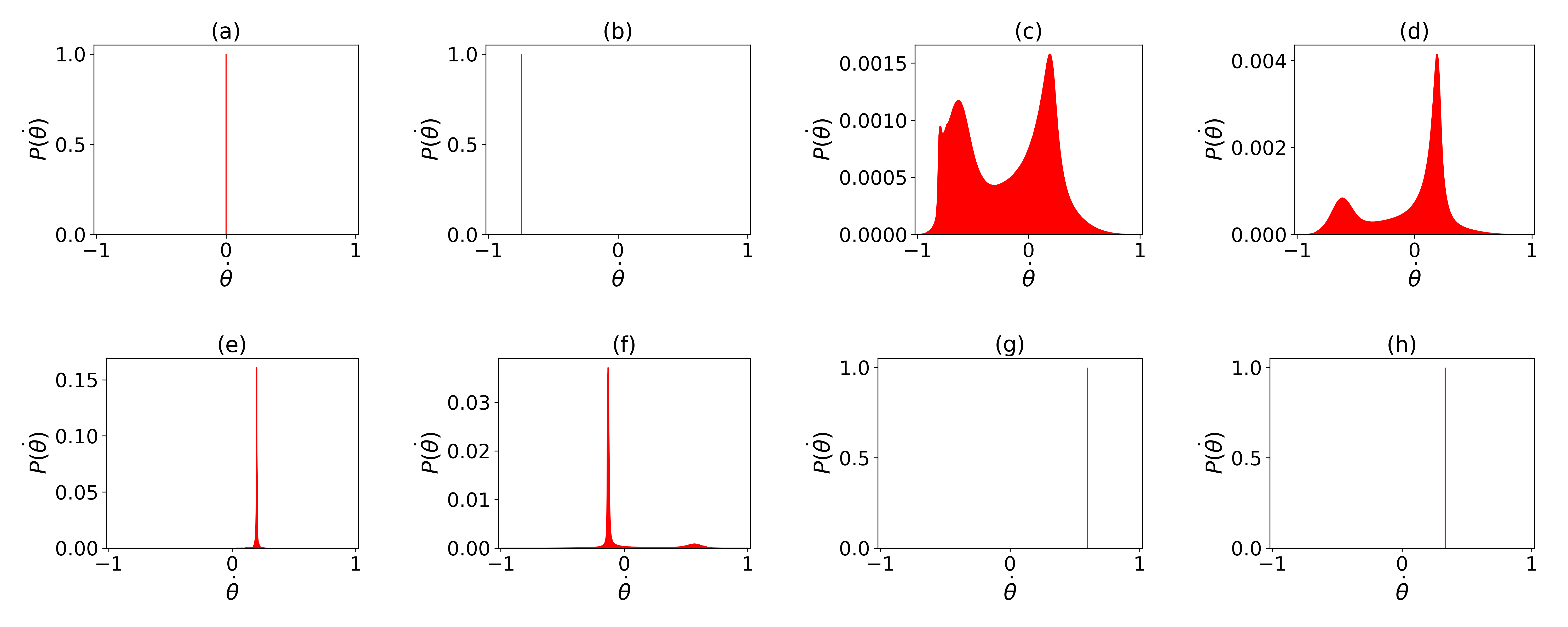}
\caption{(Color online) The probability density function of the angular velocity, $p(\dot\theta)$, for $\omega_0\tau=$ a) $0$, b) $1.6$, c) $1.9$, d) $2.04$, e) $2.2$, f) $3.92$, g) $4.08$, h) $4.9$. The size of the network is $N=1000$, its degree is $k=10$ and the intrinsic angular velocity of the oscillators is $\omega_0 = \pi/2$.}
\label{fig3:Pw}
\end{figure*}
%%%%%%%%%%%%%%%%%%%%%%%%%%%%%%%%%%%%%%%%%
%%%%%%%%%%%%%%%%%%%%%%%%%%%%%%%%%%%%%%%%Fig4
%\begin{figure*}[!b]
%\centering
%\includegraphics[width=\textwidth]{newfigs/pw25n}
%\caption{(Color online) The probability density function of the frequencies in a regular network with $N=1000$ for different phase shift values $\omega_0\tau$ . The value of $\omega_0\tau$ is written on top of each panel, with $\omega_0 = \pi/2=1.57$. $\omega_0\tau=$ a) $0$, b) $1.6$, c) $1.9$, d) $2.04$, e) $2.2$, f) $3.92$, g) $4.08$, h) $4.9$.}
%\label{fig4:Pw25}
%\end{figure*}
%%%%%%%%%%%%%%%%%%%%%%%%%%%%%%%%%%%%%%%%%
%%%%%%%%%%%%%%%%%%%%%%%%%%%%%%%%%%%%%%%%%%  Fig4
\begin{figure*}[]
\includegraphics[width=\textwidth]{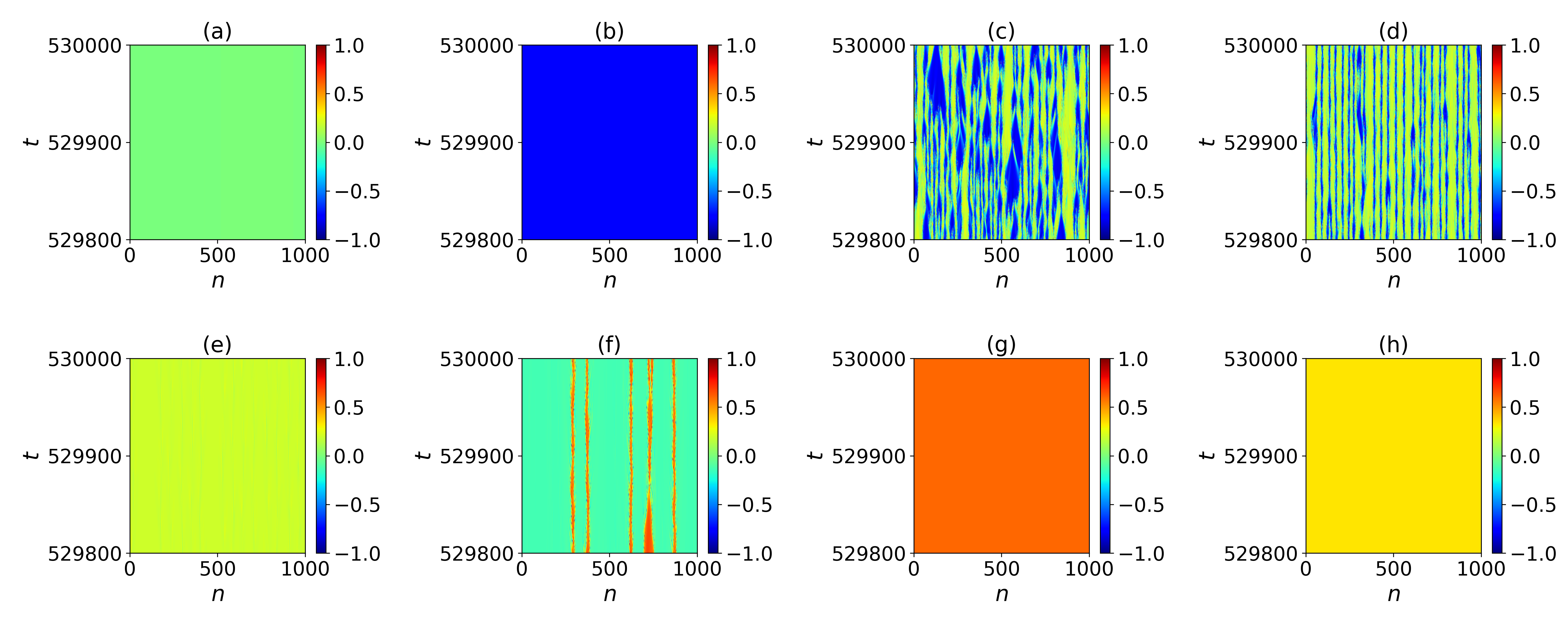}
\caption{Time evolution of angular velocity of the oscillators in a time window for $\omega_0\tau=$ a) $0$, b) $1.6$, c) $1.9$, d) $2.04$, e) $2.2$, f) $3.92$, g) $4.08$, h) $4.9$. The size of the network is $N=1000$, its degree is $k=10$ and the intrinsic angular velocity  of the oscillators is $\omega_0 = \pi/2$.}
\label{fig4:Ft}
\end{figure*}
%%%%%%%%%%%%%%%%%%%%%%%%%%%%%%%%%%%%%%%%
%%%%%%%%%%%%%%%%%%%%%%%%%%%%%%%%%%%%%%%%%%  Fig5
%\begin{figure*}[!t]
%\includegraphics[width=\textwidth]{newfigs/freq25n}
%\caption{Time evolution of frequencies of the oscillators in a time window for  $\omega_0 = \pi/2=1.57$.$\omega_0\tau=$ a) $0$, b) $1.6$, c) $1.9$, d) $2.04$, e) $2.2$, f) $3.92$, g) $4.08$, h) $4.9$.}
%\label{fig5:Ft25}
%\end{figure*}
%%%%%%%%%%%%%%%%%%%%%%%%%%%%%%%%%%%%%%%%
%%%%%%%%%%%%%%%%%%%%%%%%%%%%%%%%%%%%%%Fig5
\begin{figure*}[]
\includegraphics[width=\textwidth]{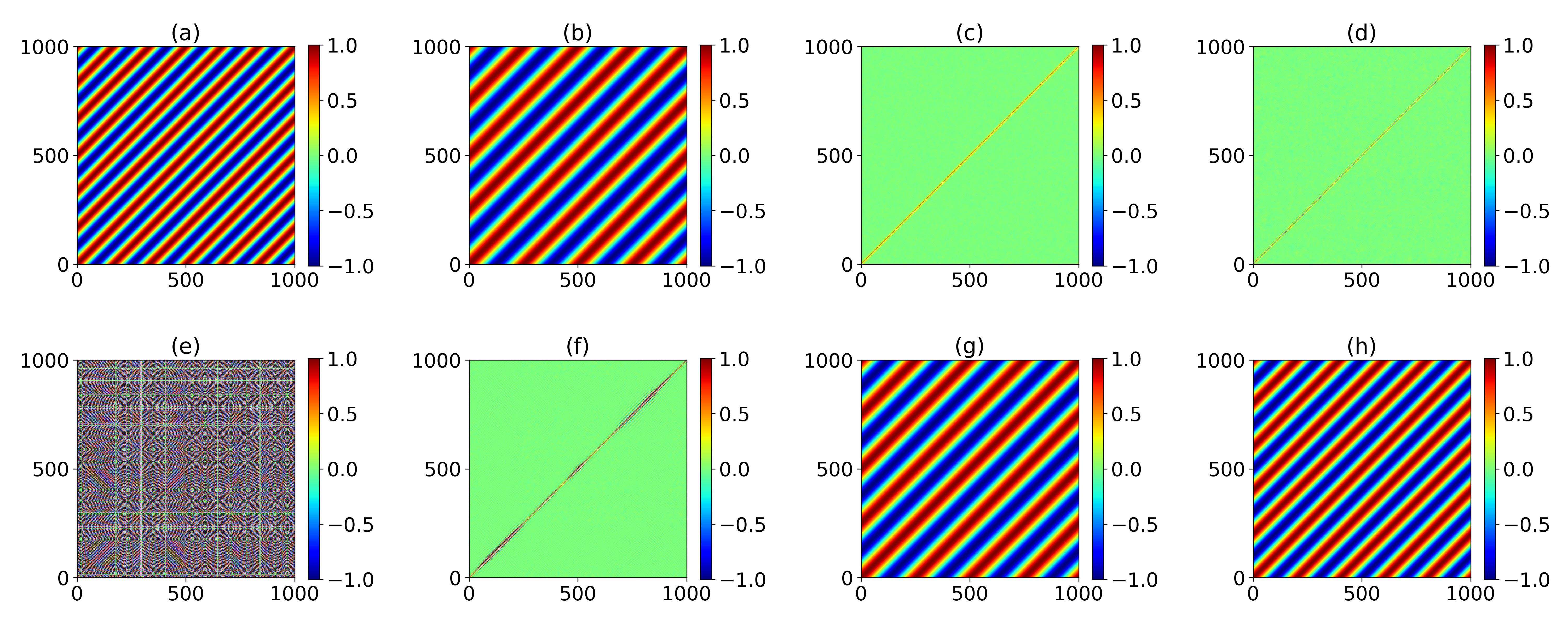}
\caption{(Color online) Density plot of the correlation matrix elements for  $\omega_0\tau=$ a) $0$, b) $1.6$, c) $1.9$, d) $2.04$, e) $2.2$, f) $3.92$, g) $4.08$, h) $4.9$. The size of the network is $N=1000$, its degree is $k=10$ and the intrinsic angular velocity of the oscillators is $\omega_0 = \pi/2$.}
\label{fig5:C}
\end{figure*}
%%%%%%%%%%%%%%%%%%%%%%%%%%%%%%%%%%%%%%%%
%%%%%%%%%%%%%%%%%%%%%%%%%%%%%%%%%%%%%%%%%Fig6
\begin{figure*}
\centering
\includegraphics[width=\columnwidth]{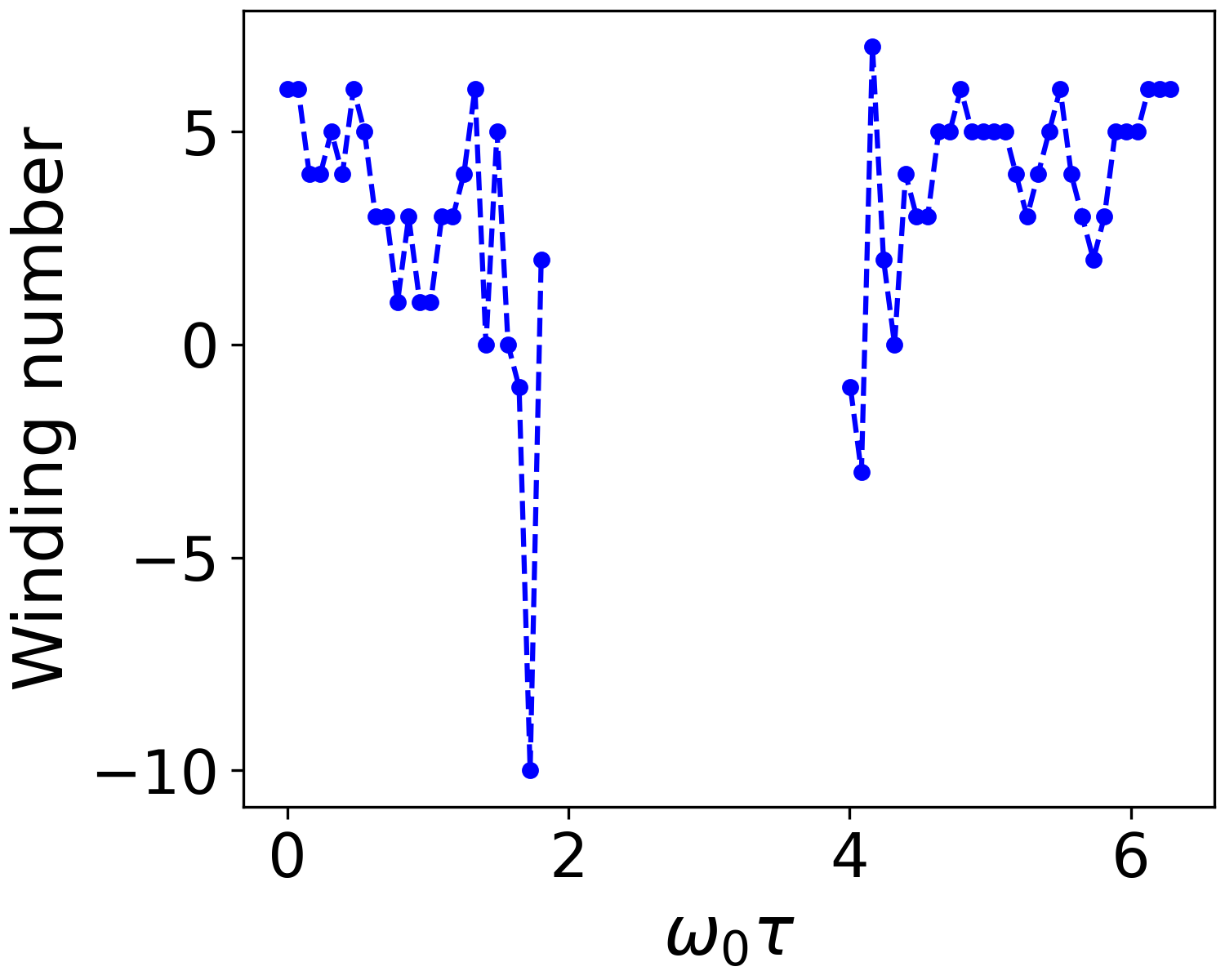}
\includegraphics[width=\columnwidth]{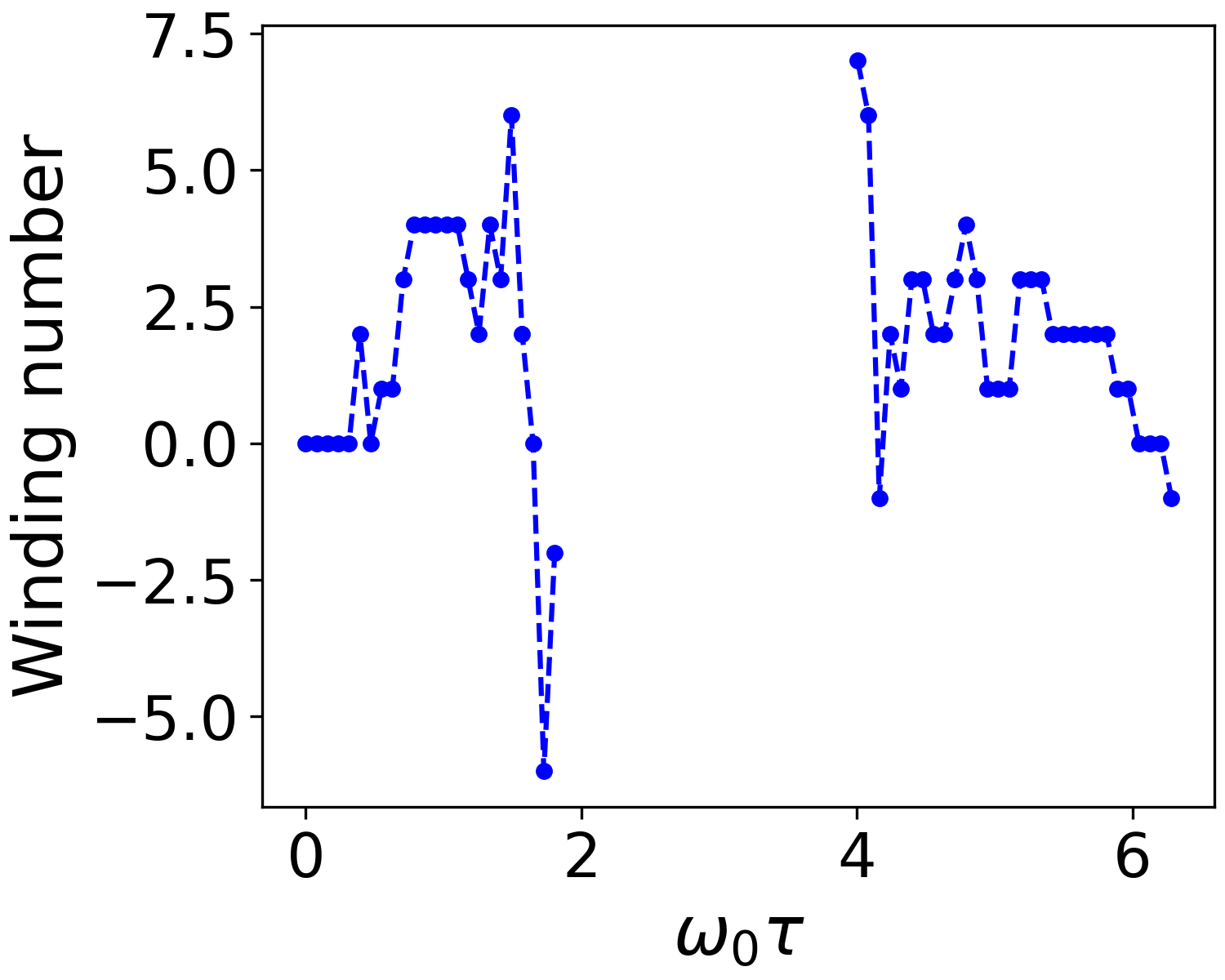}
\caption{Winding number of regular helical states versus the phase shift $\omega_{0}\tau$, for two different initial phase distributions (a) non-zero widing number with $r=0$ for $\tau=0$, and (b) zero widing number with $r=1$ for $\tau=0$. The size of the network is $N=1000$, its degree is $k=10$ and the intrinsic angular velocity of the oscillators is $\omega_0 = \pi/2$.}
% Note that in the interval between two and four the winding number is not defined.}
\label{fig6:Wi}
\end{figure*} 
%%%%%%%%%%%%%%%%%%%%%%%%%%%%%%%%%%%%
%%%%%%%%%%%%%%%%%%%%%%%%%%%%%%%%%%%%%%Fig6
%\begin{figure*}[!t]
%\includegraphics[width=\textwidth]{newfigs/d25n}
%\caption{(Color online) Density plot of the correlation matrix elements in a regular network with $N = 1000$ and k = 10 . $\omega_0 = \pi/2=1.57$. $\omega_0\tau=$ a) $0$, b) $1.6$, c) $1.9$, d) $2.04$, e) $2.2$, f) $3.92$, g) $4.08$, h) $4.9$.}
%\label{fig6:C25}
%\end{figure*}
%%%%%%%%%%%%%%%%%%%%%%%%%%%%%%%%%%%%%%%%
%%%%%%%%%%%%%%%%%%%%%%%%%%%%%%%%%%%Fig7
\begin{figure*}
\includegraphics[width=\textwidth]{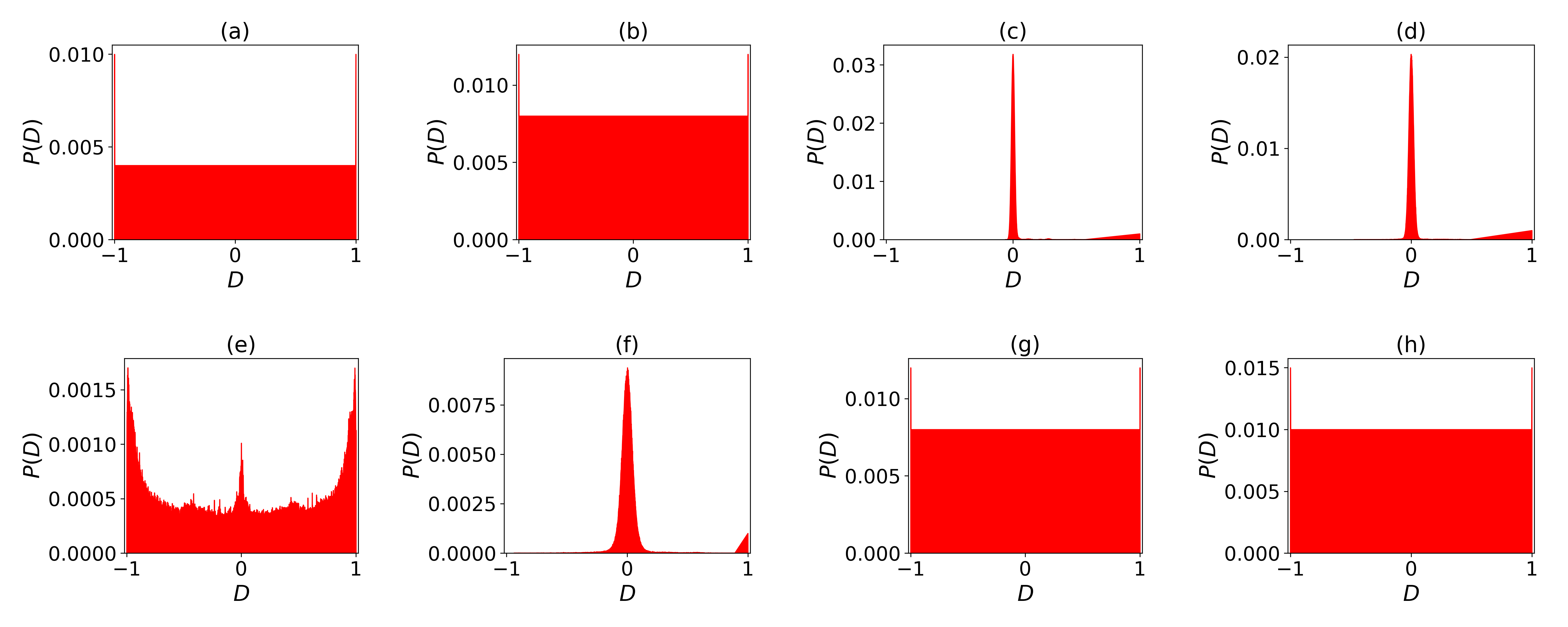}
\caption{(Color online) Probability density of of the correlation matrix elements for $\omega_0\tau=$ a) $0$, b) $1.6$, c) $1.9$, d) $2.04$, e) $2.2$, f) $3.92$, g) $4.08$, h) $4.9$. The size of the network is $N=1000$, its degree is $k=10$ and the intrinsic angular velocity of the oscillators is $\omega_0 = \pi/2$.}
\label{fig7:PD}
\end{figure*}
%%%%%%%%%%%%%%%%%%%%%%%%%%%%%%%%%%%%

%%%%%%%%%%%%%%%%%%%%%%%%%%%%%%%%%%%Fig11
%\begin{figure*}
%\includegraphics[width=\textwidth]{newfigs/pd25n}
%\caption{(Color online) Probability density of of the correlation matrix elements. $\omega_0 = \pi/2=1.57$. $\omega_0\tau=$ a) $0$, b) $1.6$, c) $1.9$, d) $2.04$, e) $2.2$, f) $3.92$, g) $4.08$, h) $4.9$.}
%\label{fig11:PD25}
%\end{figure*}
%%%%%%%%%%%%%%%%%%%%%%%%%%%%%%%%%%%%
%%%%%%%%%%%%%%%%%%%%%%%%%%%%%%%%%%%%%Fig8
\begin{figure*}
\includegraphics[width=\columnwidth]{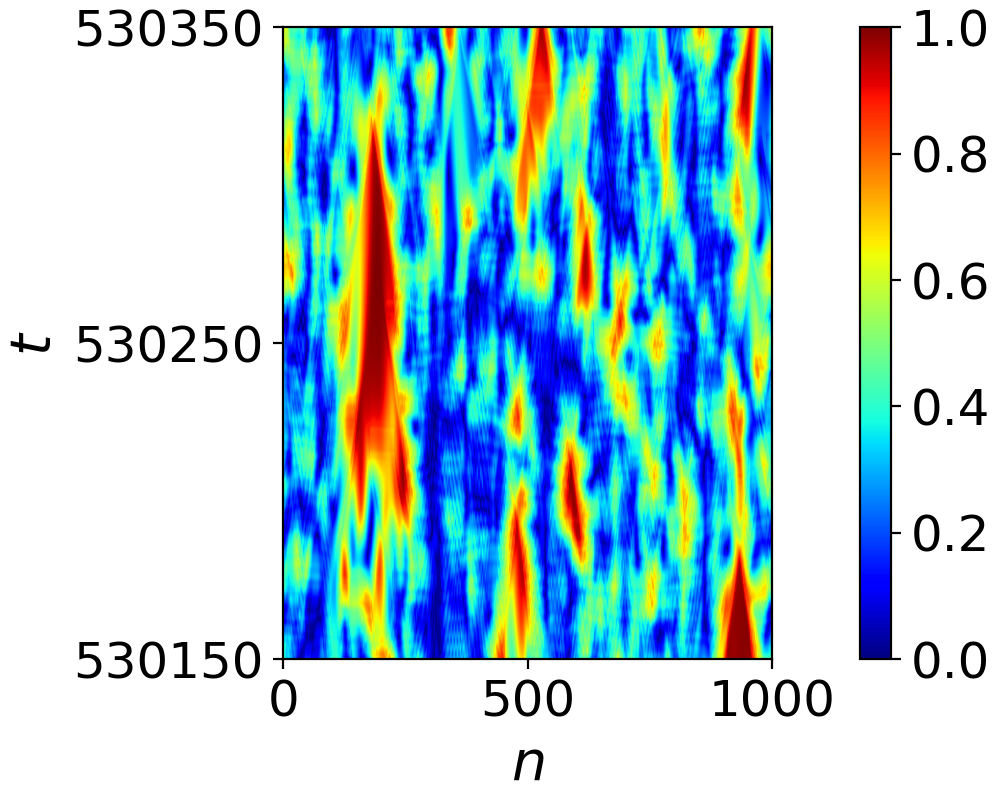}
\includegraphics[width=\columnwidth]{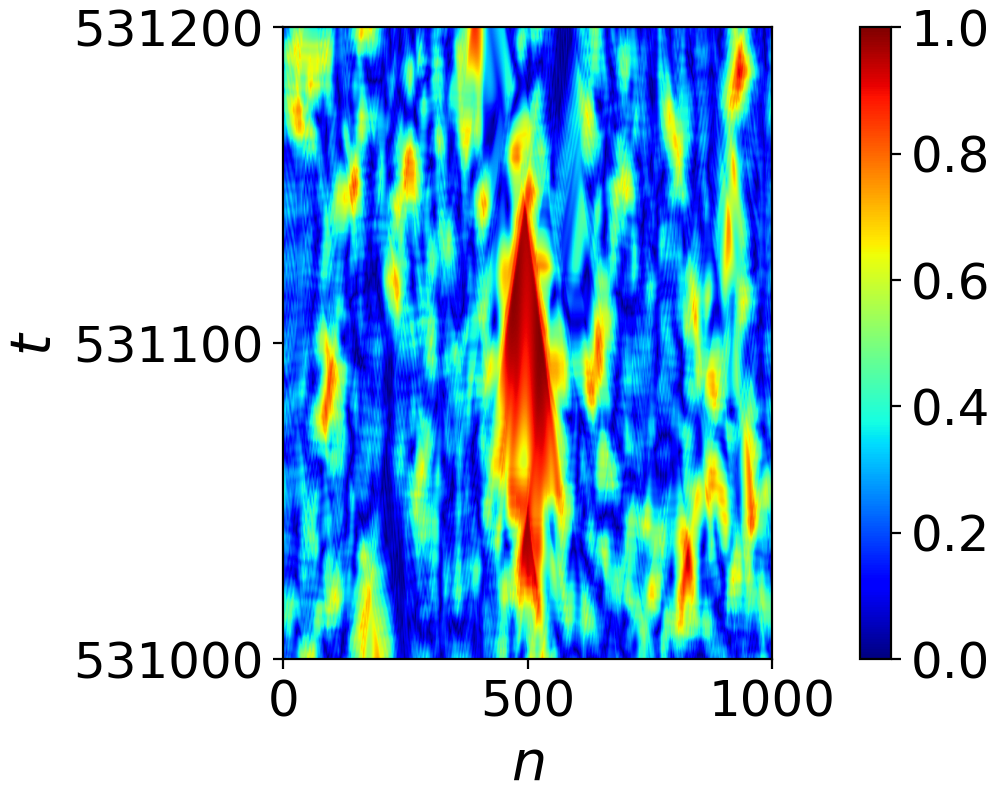}
\caption{ (Color online)Temporal evolution of the local order parameter with $50$ neighbors for (a) $\omega_0\tau=1.84$, and (b) $1.85$. The size of the network is $N=1000$, its degree is $k=10$ and the intrinsic angular velocity of the oscillators is $\omega_0 = \pi/2$.}
\label{fig8:Ch}
\end{figure*}
%%%%%%%%%%%%%%%%%%%%%%%%%%%%%%%%%%%%
%%%%%%%%%%%%%%%%%%%%%%%%%%%%%%%%%%%%%Fig9
\begin{figure*}
\includegraphics[width=\columnwidth]{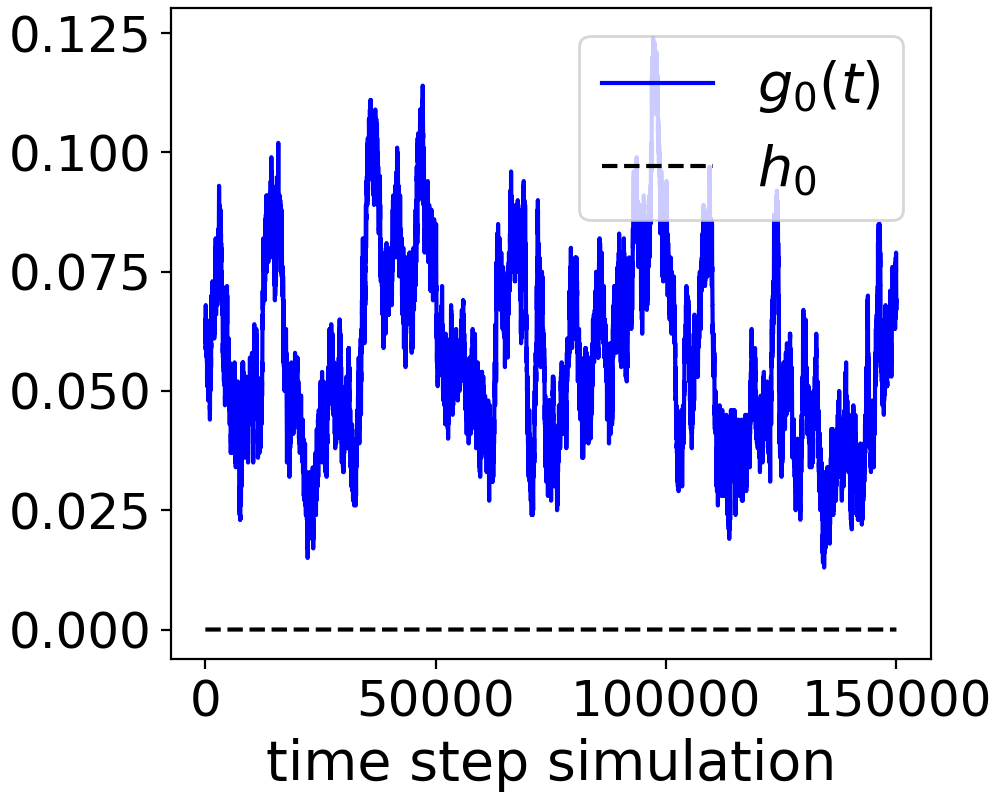}
\includegraphics[width=\columnwidth]{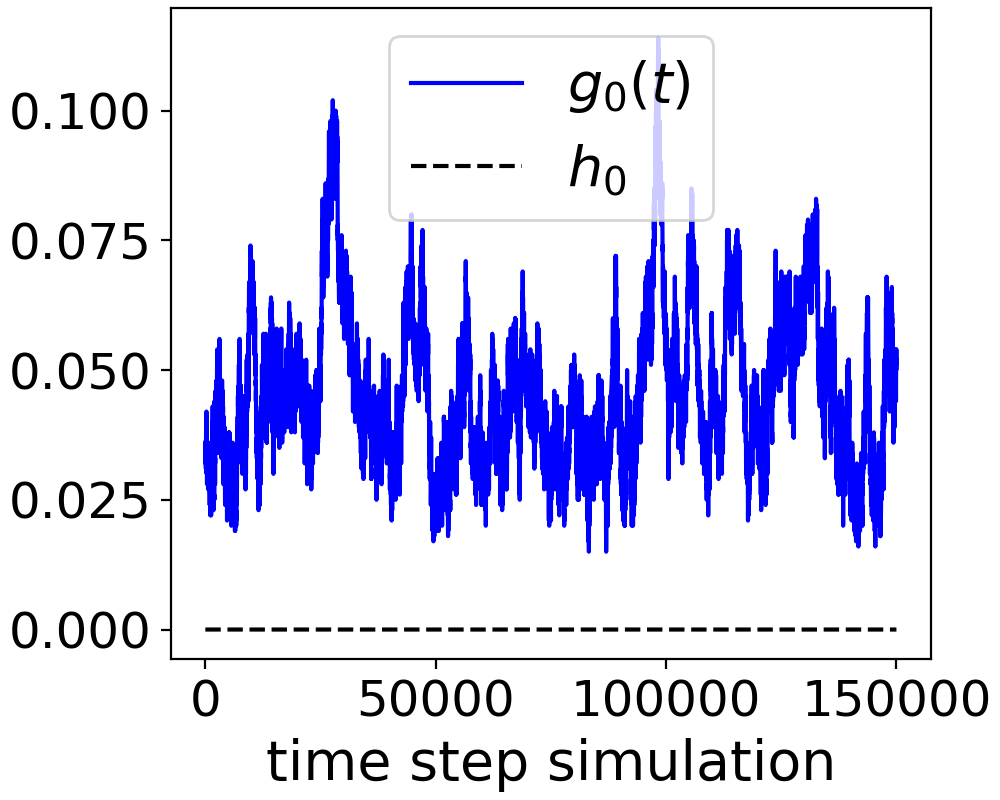}
\caption{(Color online) $g_0(t)$ versus time step sumulation after reaching the stationary state and $h_0$ for (a): $\tau=1.84$, and (b): $\tau=1.85$. }
\label{fig9:Ch2}
\end{figure*}
%%%%%%%%%%%%%%%%%%%%%%%%%%%%%%%%%%%%%%%%
%%%%%%%%%%%%%%%%%%%%%%%%%%%%%%%%%%%%%%%%%Fig13
%\begin{figure*}
%\includegraphics[scale=0.073]{newfigs/phase29n}
%\caption{Time evolution of phases of the oscillators in  a time window for  $\omega_0 = \pi/2=1.57$. $\omega_0\tau=$ a) $0$, b) $1.6$, c) $1.9$, d) $2.04$, e) $2.2$, f) $3.92$, g) $4.08$, h) $4.9$.}
%\label{Pt}
%\end{figure*}
%%%%%%%%%%%%%%%%%%%%%%%%%%%%%%%%%%%%%%%%%%
%%%%%%%%%%%%%%%%%%%%%%%%%%%%%%%%%%%%%%%%%Fig14
%\begin{figure*}
%\includegraphics[scale=0.073]{newfigs/phase25n}
%\caption{Time evolution of phases of the oscillators in  a time window for  $\omega_0 = \pi/2=1.57$. $\omega_0\tau=$ a) $0$, b) $1.6$, c) $1.9$, d) $2.04$, e) $2.2$, f) $3.92$, g) $4.08$, h) $4.9$.}
%\label{Pt25}
%\end{figure*}
%%%%%%%%%%%%%%%%%%%%%%%%%%%%%%%%%%%%%%%%%%

\subsection{Numerical simulations}

Here we represent and discuss the results obtained by the numerical simulations. We used the fourth-order Runge-Kutta method for integrating the differential equations Eq.~\eqref{mainEq2} with integrating time step $dt=0.01$. Also, the results obtained from the fourth-order Runge-Kutta method are confirmed by the sixth-order predictor-corrector method. The initial phase of the oscillators is taken randomly from a uniform phase distribution in the interval $[0, 2\pi]$. The oscillators evolved independently up to the time interval $[0, \tau]$, with their interaction commencing at $t>\tau$.

Fig.~\ref{fig2:rw}-(a), (b) illustrates the average collective angular velocity $\Omega$, and the long-time averaged order parameter, $r_{\infty}$,  as defined by Eqs.\ref{Omega} and \ref{rinf} respectively.  The blue dots represent the results corresponding to 100 realizations starting from distinct initial phase distributions. 
Fig.~\ref{fig2:rw}-(a), presents oscillators' long-time averaged collective angular velocity in the rotating frame as a function of 
phase shift $\omega_0\tau$. We observe successive patterns of  
decreased average angular velocity as a function of $\omega_0\tau$  and rapid jumps at certain values of $\omega_0\tau$. This is comparable to the variation of the angular velocity of the fully synchronized state versus the phase shift $\omega_0\tau$, shown in Fig.~\ref{fig1:analytic}. 

Fig.~\ref{fig2:rw}-(b) provides a measure of global synchrony  ($r_{\infty}$) after the system has reached its stationary state. The result indicates that, for $0<\omega_0\tau\lesssim 2.1$ and $3.9\lesssim\omega_0\tau\lesssim 6.1$, the system can reach a fully synchronized state 
($r_{\infty}=1$) or states with $r_{\infty}=0$, which we will show that these are regular phase-locked states with constant phase lags.  

However, for $2.1 \lesssim\omega_0\tau\lesssim 3.9$, the order parameter is exactly zero for all the initial phase distributions. In what follows, we show these states are random phase-locked or {\em glassy} states.  The synchrony-forbidden region has attracted significant attention due to its unique characteristics. Yeung and Strogatz employed stability analysis in the infinite-N limit, reformulating the model as a Fokker-Planck equation to identify the forbidden synchrony regions\cite{yeung1999time}, which aligns with our numerical findings in Fig.\ref{fig2:rw}. The origin of this behavior will be discussed in detail in subsequent sections.

To gain insight into the local dynamics, we present the probability density function and time evolution of the angular velocity of oscillators for a specific initial phase distribution in Figs. ~\ref{fig3:Pw} and \ref{fig4:Ft}. The panels (a), (b), (e), (g), and (h) in both figures demonstrate that, for the time delays corresponding to the collective branches, shown in  Fig. ~\ref{fig2:rw}-(a), the angular velocity of all oscillators reaches a constant value in the stationary state, which depends on the value of time delay. This indicates that the system remains frequency-locked, and delay causes a negative shift in collective angular velocity, $\Omega$, in each branch. 

However, the angular velocity probability density functions  in the transition regions between different branches are bimodal,  as shown in 
Figs. ~\ref{fig3:Pw}-(c), (d), and (f), indicating the coexistence of two major frequencies which are higher and lower than the intrinsic angular velocity of oscillators. The time evolution plots of the angular velocity of oscillators, illustrated in Fig.~\ref{fig4:Ft}-(c), (d), and (f), confirm that while in some cases there is a coexistence between two groups of oscillators with different angular velocities, however during the dynamics, the angular velocity of some oscillators transfer between the two dominant values corresponding to the peaks of angular velocity distributions. Therefore, within the transition regions, one expects no phase coherence among the oscillators.

Fig.~ \ref{fig5:C}, illustrates the time-averaged correlation matrices ($D$) given by Eq.\ref{D}. The x and y axes represent the number of oscillators, and the colors display the heat map of the values of the correlation matrix. 
In the zero time delay condition, there are periodic stripe patterns of in-phase (red)
and antiphase (blue) which correspond to the helical patterns that have been previously explored in 
regular networks~\cite{wiley2006size, esfahani2012noise}. These regular helical patterns exist for the time delays corresponding to the first and third branches of Fig.~\ref{fig2:rw} (pnales (a), (b), (g), and (h)). However, for the middle branch, the correlation matrix shows no order which is due to the random freezing of phase variables. At the transition points the correlation is zero for all pairs of oscillators indicating the absence of phase coherence in the system.

Regular helical patterns are topological and can be assigned a winding number. Considering  $\Delta_{i,i+1} = \theta_i - \theta_{i+1}$, the successive angle differences modulo $2\pi$ within the interval $(-\pi, \pi]$, we find that the sum over the ring of $\Delta_{i,i+1}$ results in an integer multiple of $2\pi$. This  leads us to the definition of the integer winding number $q$

\begin{equation}
q=\frac{1}{2\pi} \sum_{i=1}^{N} \Delta_{i,i+1}. 
\end{equation}

Calculation of the widening number, nables us to explore how the non-zero time delay modifies the helical patterns. 
 Figs~\ref{fig6:Wi}-{a} and \ref{fig6:Wi}-{b} present the variation of the steady-state winding number of the regular patterns versus time delay, for two realizations with distinct initial phase distributions. Both panels show that the winding number changes irregularly when the time delay increases.

In Fig.~\ref{fig7:PD}, we present the probability density function of the time-averaged pair correlation matrix elements, denoted as $p(D)$. As expected, for the regular helical patterns, this figure illustrates a uniform distribution ranging from -1 to 1 for the elements of the pair correlation function depicted in Fig.~\ref{fig7:PD}-(a), (b), (g), and (h).

For $\omega\tau=2.2$, which belongs to the middle branch of Fig.~\ref{fig2:rw}, it is observed that $p(D)$ exhibits a U-shaped distribution (Fig.~\ref{fig7:PD}-(e)). This characteristic signifies a uniform phase distribution, implying that the phases of the oscillators are randomly frozen within the interval $(-\pi, \pi]$. The sharp peak centered at zero denotes the incoherent dynamics of a small group of oscillators. Therefore, for this specific time delay, there exists a coexistence of a glassy state and incoherent dynamics. 

At the transition points, the distribution $p(D)$ exhibits bimodal characteristics, with one peak centered at $0$ and the other at $+1$ (refer to Fig.~\ref{fig7:PD}-(c), (d), (f)). The prominent Gaussian peaks centered at zero signify the incoherent dynamics of oscillators, while the smaller peaks at $+1$ indicate a limited group of synchronized oscillators. The Gaussian peak at zero arises from the central limit theorem due to the absence of temporal correlation in the phase differences of pairs of oscillators. The coexistence of synchrony and incoherent dynamics serves as an indicator of chimera states at the transition points. Such chimera states have been previously observed in the time-delayed Kuramoto model within small-world networks~\cite{ameli2021time}. In the subsequent subsections, we shall investigate, in detail, the nature of these chimera states and establish their classification.

\subsection{Chimera states}

Fig.~\ref{fig8:Ch} presents the temporal variation of the local order parameter as defined by Equation~\eqref{r-local} for $m=50$ neighbors within a time window spanning from $531000$ to $531200$ for $\omega_0\tau=1.84, 1.85$ at the transition zone. The red regions illustrate a group of oscillators exhibiting high synchrony. In both instances, one observes the coexistence of coherent and incoherent groups of phase oscillators, indicating the presence of Chimera states.  We employ the method proposed by Kemeth et al.~\cite{kemeth2016classification} to quantitatively ascertain the type of these Chimera states.

We utilize two functions, $g_0(t)$ (spatial coherence of neighbors) and $h_0$ (temporal correlation of neighbors), to determine the types of chimera states. In this context, we initially calculate the distance between neighboring oscillators by measuring the magnitude of the subtraction of their phase vectors
\begin{equation}
\mathbf{S}_{i}=|e^{i\theta_i}-e^{i\theta_{i+1}}|=2|\sin\big({\frac{\theta_i-\theta_{i+1}}{2}}\big)|,~~~~~~~i = 1,..N
\label{si}
\end{equation} 
Then we obtain  the probability density function of $\mathbf{S}$, $g(t,\mathbf{S})$, from which we define $g_{0}(t)$ as
\begin{equation}
g_0(t)=\int_{0}^{\delta} g(t,\mathbf{S}) d\mathbf{S},
\label{g0t}
\end{equation} 
 where $\delta=0.01\max{\mathbf{S}}$. It is clear that $0 \leq g_0(t) \leq 1$ at time $t$. In the case of a large difference between neighboring phase vectors, $g_{0}(t) \approx 0$, and when all neighboring phase vectors are very close to each other, $g_{0}(t) \approx 1$. 
 
We use the correlation matrix $D$, defined by Eq.~\eqref{D}, to calculate the temporal correlation of neighboring oscillators.
 Defining $\rho_i$ as 
\begin{equation}
\mathbf{\rho}_i=D_{i,i+1},~~~~~~~i = 1,...,N
\label{rhoi}
\end{equation} 
we can calculate $h_0$ as follows
\begin{equation}
h_0=\int_{\gamma}^{1} h(|\mathbf{\rho}|) d|\mathbf{\rho}|,
\label{h0}
\end{equation} 
in which $\gamma=0.99$ and $h(|\mathbf{\rho}|)$ is the probability density function of absolute value of $\rho_i$'s. 

%Our method is similar to Kemeth et al ~\cite{kemeth2016classification}. $\{\forall t: 0<g_0(t)<1\}$ denotes chimera state. The constancy, regular oscillatory, and irregular oscillatory of $g_0(t)$ in time denote stationary, breathing, and turbulent chimera states, respectively. 
%Transient chimera is identified by $\{\{\exists t_0: g_0(t)=0 \lor g_0(t)=1\} \land \{\forall t<t_0: 0<g_0(t)<1\}\}$. $h_0\approx0$ and $h_0>0$ also represent moving and static states.

Fig.~\ref{fig9:Ch2} demonstrates that for both $\omega_0\tau=1.84$ and $1.85$, the value of $h_0$ is equal to zero, and $g_0(t)$ fluctuates irregularly within a limited range between zero and one at the stationary state. Kemeth et al.~\cite{kemeth2016classification} discuss that these behaviors of $h_0$ and $g_0(t)$ indicate a moving turbulent chimera state.

\section{Conclusion}
\label{conclusion}
In summary, we examined the impact of time-delayed coupling on the synchronization of the Kuramoto model within regular networks, specifically emphasizing both global and local synchronization dynamics. Our investigation yields several key insights into how time delays affect the collective behavior of oscillators in networks characterized by high clustering coefficients.

In line with earlier findings from small-world networks~\cite{ameli2021time}, our current analysis identified distinct dynamical regimes based on the phase shift $\omega_0\tau$. When $\cos(\omega_0\tau) > 0$, i.e. for the intervals $0 \leq \omega_0\tau \leq \frac{\pi}{2}$ and $\frac{3\pi}{2} \leq \omega_0\tau \leq 2\pi$, we observe multistability among different topological states, each characterized by an integer winding numbers. A winding number of $0$ corresponds to full synchrony, represented by $r=1$, while nonzero winding numbers indicate helical phase-locked states, which result in $r=0$.

Conversely, when $\cos(\omega_0\tau) < 0$, we encounter a region that is considered {\em forbidden}, specifically the interval 
$2 \lesssim \omega_{0}\tau \lesssim 4$, Within this range, the system cannot achieve synchronization, resulting instead in a randomly quenched phase, often referred to as a {\em glassy state}, or a coexistence of the glassy state and incoherent dynamics. These observations align with predictions from stability analyses conducted in the infinite-N limit, where the system exhibits a clear transition between synchronized and desynchronized states~\cite{yeung1999time}. 

The above regimes correspond to three distinct branches of angular velocity in which all oscillators rotate with the same angular velocity. In each branch, the angular velocity of the oscillators decreases as the time delay increases.  

In the narrow transition intervals between these branches, specifically in the ranges $\frac{\pi}{2} \lesssim \omega_{0}\tau \lesssim 2 $
 and $4 \lesssim \omega_{0}\tau \lesssim \frac{3\pi}{2} $, we observe a significant number of dynamically incoherent oscillators. Within this group, some smaller clusters of coherent oscillators form and dissipate over finite time intervals. The distribution of angular velocities in these transition regions is bimodal, indicating that the oscillators rotate mostly at two dominant angular velocities. We have demonstrated that these configurations represent Chimera states and established that their classes are  moving turbulence.

The results of this study provide valuable insights into the complex dynamics of synchronization in regular networks under the influence of time delays. The identification of synchronization "forbidden zones" and the detailed exploration of local dynamics underscore the critical role that time delay plays in determining the collective behavior of the system. These findings possess broader implications for understanding synchronization in various real-world systems, including neuronal networks, power grids, and communication systems, wherein time delays are inherent due to finite transmission speeds.

%Furthermore, the interplay between network topology and time delay offers new avenues for controlling synchronization in engineered systems. 

Overall, this study contributes significantly to the expanding body of knowledge regarding synchronization in complex networks, underscoring the necessity for further research into the interplay between different topologies and coupling functions in conjunction with time delays. Future endeavors could broaden this analysis to encompass more heterogeneous networks, examine the impacts of non-uniform time delays, or investigate the robustness of synchronization in the face of varying external perturbations. By persistently deepening our understanding of these dynamics, we can more effectively harness the potential of synchronization in both natural and engineered systems. Through the adjustment of network structures or time delays, it may be feasible to achieve desired synchronization states or avert unwanted desynchronization. This consideration could be particularly pertinent in applications where the maintenance of a synchronized state is imperative, such as in the coordination of distributed systems or the stabilization of power networks.

%Although connections can be weighted by the dynamics of the system \cite{ameli2022low}, here we consider a $0$ or $1$ connection, weighted by the coupling strength. Further research can investigate the effect of delay in a system with nonhomogeneous interactions.
%This comprehensive study enhances the understanding of synchronization mechanisms in regular networks and provides valuable insights into the effects of time delay on collective dynamics.

\section*{References}
%\longbibliography{cas}
\bibliography{cas}

\begin{thebibliography}{34}%
\makeatletter
\providecommand \@ifxundefined [1]{%
 \@ifx{#1\undefined}
}%
\providecommand \@ifnum [1]{%
 \ifnum #1\expandafter \@firstoftwo
 \else \expandafter \@secondoftwo
 \fi
}%
\providecommand \@ifx [1]{%
 \ifx #1\expandafter \@firstoftwo
 \else \expandafter \@secondoftwo
 \fi
}%
\providecommand \natexlab [1]{#1}%
\providecommand \enquote  [1]{``#1''}%
\providecommand \bibnamefont  [1]{#1}%
\providecommand \bibfnamefont [1]{#1}%
\providecommand \citenamefont [1]{#1}%
\providecommand \href@noop [0]{\@secondoftwo}%
\providecommand \href [0]{\begingroup \@sanitize@url \@href}%
\providecommand \@href[1]{\@@startlink{#1}\@@href}%
\providecommand \@@href[1]{\endgroup#1\@@endlink}%
\providecommand \@sanitize@url [0]{\catcode `\\12\catcode `\$12\catcode
  `\&12\catcode `\#12\catcode `\^12\catcode `\_12\catcode `\%12\relax}%
\providecommand \@@startlink[1]{}%
\providecommand \@@endlink[0]{}%
\providecommand \url  [0]{\begingroup\@sanitize@url \@url }%
\providecommand \@url [1]{\endgroup\@href {#1}{\urlprefix }}%
\providecommand \urlprefix  [0]{URL }%
\providecommand \Eprint [0]{\href }%
\providecommand \doibase [0]{http://dx.doi.org/}%
\providecommand \selectlanguage [0]{\@gobble}%
\providecommand \bibinfo  [0]{\@secondoftwo}%
\providecommand \bibfield  [0]{\@secondoftwo}%
\providecommand \translation [1]{[#1]}%
\providecommand \BibitemOpen [0]{}%
\providecommand \bibitemStop [0]{}%
\providecommand \bibitemNoStop [0]{.\EOS\space}%
\providecommand \EOS [0]{\spacefactor3000\relax}%
\providecommand \BibitemShut  [1]{\csname bibitem#1\endcsname}%
\let\auto@bib@innerbib\@empty
%</preamble>
\bibitem [{\citenamefont {Strogatz}(2012)}]{strogatz2012sync}%
  \BibitemOpen
  \bibfield  {author} {\bibinfo {author} {\bibfnamefont {Steven~H}\
  \bibnamefont {Strogatz}},\ }\href@noop {} {\emph {\bibinfo {title} {Sync: How
  order emerges from chaos in the universe, nature, and daily life}}}\
  (\bibinfo  {publisher} {Hachette Books},\ \bibinfo {year} {2012})\BibitemShut
  {NoStop}%
\bibitem [{\citenamefont {Nijmeijer}\ and\ \citenamefont
  {Rodriguez-Angeles}(2003)}]{nijmeijer2003synchronization}%
  \BibitemOpen
  \bibfield  {author} {\bibinfo {author} {\bibfnamefont {Henk}\ \bibnamefont
  {Nijmeijer}}\ and\ \bibinfo {author} {\bibfnamefont {Alejandro}\ \bibnamefont
  {Rodriguez-Angeles}},\ }\href@noop {} {\emph {\bibinfo {title}
  {Synchronization of mechanical systems}}},\ Vol.~\bibinfo {volume} {46}\
  (\bibinfo  {publisher} {World Scientific},\ \bibinfo {year}
  {2003})\BibitemShut {NoStop}%
\bibitem [{\citenamefont {Mirollo}\ and\ \citenamefont
  {Strogatz}(1990)}]{mirollo1990synchronization}%
  \BibitemOpen
  \bibfield  {author} {\bibinfo {author} {\bibfnamefont {Renato~E}\
  \bibnamefont {Mirollo}}\ and\ \bibinfo {author} {\bibfnamefont {Steven~H}\
  \bibnamefont {Strogatz}},\ }\bibfield  {title} {\enquote {\bibinfo {title}
  {Synchronization of pulse-coupled biological oscillators},}\ }\href@noop {}
  {\bibfield  {journal} {\bibinfo  {journal} {SIAM Journal on Applied
  Mathematics}\ }\textbf {\bibinfo {volume} {50}},\ \bibinfo {pages}
  {1645--1662} (\bibinfo {year} {1990})}\BibitemShut {NoStop}%
\bibitem [{\citenamefont {Tyson}(1973)}]{tyson1973some}%
  \BibitemOpen
  \bibfield  {author} {\bibinfo {author} {\bibfnamefont {John~J}\ \bibnamefont
  {Tyson}},\ }\bibfield  {title} {\enquote {\bibinfo {title} {Some further
  studies of nonlinear oscillations in chemical systems},}\ }\href@noop {}
  {\bibfield  {journal} {\bibinfo  {journal} {The Journal of Chemical Physics}\
  }\textbf {\bibinfo {volume} {58}},\ \bibinfo {pages} {3919--3930} (\bibinfo
  {year} {1973})}\BibitemShut {NoStop}%
\bibitem [{\citenamefont {Blasius}\ \emph {et~al.}(1999)\citenamefont
  {Blasius}, \citenamefont {Huppert},\ and\ \citenamefont
  {Stone}}]{blasius1999complex}%
  \BibitemOpen
  \bibfield  {author} {\bibinfo {author} {\bibfnamefont {Bernd}\ \bibnamefont
  {Blasius}}, \bibinfo {author} {\bibfnamefont {Amit}\ \bibnamefont {Huppert}},
  \ and\ \bibinfo {author} {\bibfnamefont {Lewi}\ \bibnamefont {Stone}},\
  }\bibfield  {title} {\enquote {\bibinfo {title} {Complex dynamics and phase
  synchronization in spatially extended ecological systems},}\ }\href@noop {}
  {\bibfield  {journal} {\bibinfo  {journal} {Nature}\ }\textbf {\bibinfo
  {volume} {399}},\ \bibinfo {pages} {354--359} (\bibinfo {year}
  {1999})}\BibitemShut {NoStop}%
\bibitem [{\citenamefont {Pikovsky}\ \emph {et~al.}(2003)\citenamefont
  {Pikovsky}, \citenamefont {Rosenblum},\ and\ \citenamefont
  {Kurths}}]{pikovsky2003synchronization}%
  \BibitemOpen
  \bibfield  {author} {\bibinfo {author} {\bibfnamefont {A}~\bibnamefont
  {Pikovsky}}, \bibinfo {author} {\bibfnamefont {M}~\bibnamefont {Rosenblum}},
  \ and\ \bibinfo {author} {\bibfnamefont {J}~\bibnamefont {Kurths}},\
  }\href@noop {} {\emph {\bibinfo {title} {Synchronization: A universal concept
  in nonlinear sciences}}}\ (\bibinfo  {publisher} {Cambridge University
  Press},\ \bibinfo {year} {2003})\BibitemShut {NoStop}%
\bibitem [{\citenamefont {Arenas}\ \emph {et~al.}(2008)\citenamefont {Arenas},
  \citenamefont {D{\'\i}az-Guilera}, \citenamefont {Kurths}, \citenamefont
  {Moreno},\ and\ \citenamefont {Zhou}}]{arenas2008synchronization}%
  \BibitemOpen
  \bibfield  {author} {\bibinfo {author} {\bibfnamefont {Alex}\ \bibnamefont
  {Arenas}}, \bibinfo {author} {\bibfnamefont {Albert}\ \bibnamefont
  {D{\'\i}az-Guilera}}, \bibinfo {author} {\bibfnamefont {Jurgen}\ \bibnamefont
  {Kurths}}, \bibinfo {author} {\bibfnamefont {Yamir}\ \bibnamefont {Moreno}},
  \ and\ \bibinfo {author} {\bibfnamefont {Changsong}\ \bibnamefont {Zhou}},\
  }\bibfield  {title} {\enquote {\bibinfo {title} {Synchronization in complex
  networks},}\ }\href@noop {} {\bibfield  {journal} {\bibinfo  {journal}
  {Physics reports}\ }\textbf {\bibinfo {volume} {469}},\ \bibinfo {pages}
  {93--153} (\bibinfo {year} {2008})}\BibitemShut {NoStop}%
\bibitem [{\citenamefont {Pecora}\ and\ \citenamefont
  {Carroll}(1998)}]{pecora1998master}%
  \BibitemOpen
  \bibfield  {author} {\bibinfo {author} {\bibfnamefont {Louis~M}\ \bibnamefont
  {Pecora}}\ and\ \bibinfo {author} {\bibfnamefont {Thomas~L}\ \bibnamefont
  {Carroll}},\ }\bibfield  {title} {\enquote {\bibinfo {title} {Master
  stability functions for synchronized coupled systems},}\ }\href@noop {}
  {\bibfield  {journal} {\bibinfo  {journal} {Physical review letters}\
  }\textbf {\bibinfo {volume} {80}},\ \bibinfo {pages} {2109} (\bibinfo {year}
  {1998})}\BibitemShut {NoStop}%
\bibitem [{\citenamefont {Nakao}\ \emph {et~al.}(2007)\citenamefont {Nakao},
  \citenamefont {Arai},\ and\ \citenamefont {Kawamura}}]{nakao2007noise}%
  \BibitemOpen
  \bibfield  {author} {\bibinfo {author} {\bibfnamefont {Hiroya}\ \bibnamefont
  {Nakao}}, \bibinfo {author} {\bibfnamefont {Kensuke}\ \bibnamefont {Arai}}, \
  and\ \bibinfo {author} {\bibfnamefont {Yoji}\ \bibnamefont {Kawamura}},\
  }\bibfield  {title} {\enquote {\bibinfo {title} {Noise-induced
  synchronization and clustering in ensembles of uncoupled limit-cycle
  oscillators},}\ }\href@noop {} {\bibfield  {journal} {\bibinfo  {journal}
  {Physical review letters}\ }\textbf {\bibinfo {volume} {98}},\ \bibinfo
  {pages} {184101} (\bibinfo {year} {2007})}\BibitemShut {NoStop}%
\bibitem [{\citenamefont {Schuster}\ and\ \citenamefont
  {Wagner}(1989)}]{schuster1989mutual}%
  \BibitemOpen
  \bibfield  {author} {\bibinfo {author} {\bibfnamefont {Heinz~Georg}\
  \bibnamefont {Schuster}}\ and\ \bibinfo {author} {\bibfnamefont {Peter}\
  \bibnamefont {Wagner}},\ }\bibfield  {title} {\enquote {\bibinfo {title}
  {Mutual entrainment of two limit cycle oscillators with time delayed
  coupling},}\ }\href@noop {} {\bibfield  {journal} {\bibinfo  {journal}
  {Progress of Theoretical Physics}\ }\textbf {\bibinfo {volume} {81}},\
  \bibinfo {pages} {939--945} (\bibinfo {year} {1989})}\BibitemShut {NoStop}%
\bibitem [{\citenamefont {Mahdavi}\ \emph {et~al.}(2025)\citenamefont
  {Mahdavi}, \citenamefont {Zarei},\ and\ \citenamefont
  {Shahbazi}}]{mahdavi2025synchronization}%
  \BibitemOpen
  \bibfield  {author} {\bibinfo {author} {\bibfnamefont {Esmaeil}\ \bibnamefont
  {Mahdavi}}, \bibinfo {author} {\bibfnamefont {Mina}\ \bibnamefont {Zarei}}, \
  and\ \bibinfo {author} {\bibfnamefont {Farhad}\ \bibnamefont {Shahbazi}},\
  }\bibfield  {title} {\enquote {\bibinfo {title} {Synchronization of two
  coupled massive oscillators in the time-delayed kuramoto model},}\
  }\href@noop {} {\bibfield  {journal} {\bibinfo  {journal} {Chaos: An
  Interdisciplinary Journal of Nonlinear Science}\ }\textbf {\bibinfo {volume}
  {35}} (\bibinfo {year} {2025})}\BibitemShut {NoStop}%
\bibitem [{\citenamefont {Ameli}\ and\ \citenamefont
  {Samani}(2024)}]{ameli2024two}%
  \BibitemOpen
  \bibfield  {author} {\bibinfo {author} {\bibfnamefont {Sara}\ \bibnamefont
  {Ameli}}\ and\ \bibinfo {author} {\bibfnamefont {Keivan~Aghababaei}\
  \bibnamefont {Samani}},\ }\bibfield  {title} {\enquote {\bibinfo {title}
  {Two-step and explosive synchronization in frequency-weighted kuramoto
  model},}\ }\href@noop {} {\bibfield  {journal} {\bibinfo  {journal} {Physica
  D: Nonlinear Phenomena}\ }\textbf {\bibinfo {volume} {470}},\ \bibinfo
  {pages} {134349} (\bibinfo {year} {2024})}\BibitemShut {NoStop}%
\bibitem [{\citenamefont {Ameli}\ and\ \citenamefont
  {Samani}(2022)}]{ameli2022low}%
  \BibitemOpen
  \bibfield  {author} {\bibinfo {author} {\bibfnamefont {Sara}\ \bibnamefont
  {Ameli}}\ and\ \bibinfo {author} {\bibfnamefont {Keivan~Aghababaei}\
  \bibnamefont {Samani}},\ }\bibfield  {title} {\enquote {\bibinfo {title}
  {Low-dimensional behavior of generalized kuramoto model},}\ }\href@noop {}
  {\bibfield  {journal} {\bibinfo  {journal} {Nonlinear dynamics}\ }\textbf
  {\bibinfo {volume} {110}},\ \bibinfo {pages} {2781--2791} (\bibinfo {year}
  {2022})}\BibitemShut {NoStop}%
\bibitem [{\citenamefont {Daniels}\ \emph {et~al.}(2003)\citenamefont
  {Daniels}, \citenamefont {Dissanayake},\ and\ \citenamefont
  {Trees}}]{daniels2003synchronization}%
  \BibitemOpen
  \bibfield  {author} {\bibinfo {author} {\bibfnamefont {BC}~\bibnamefont
  {Daniels}}, \bibinfo {author} {\bibfnamefont {STM}\ \bibnamefont
  {Dissanayake}}, \ and\ \bibinfo {author} {\bibfnamefont {BR}~\bibnamefont
  {Trees}},\ }\bibfield  {title} {\enquote {\bibinfo {title} {Synchronization
  of coupled rotators: Josephson junction ladders and the locally coupled
  kuramoto model},}\ }\href@noop {} {\bibfield  {journal} {\bibinfo  {journal}
  {Physical Review E}\ }\textbf {\bibinfo {volume} {67}},\ \bibinfo {pages}
  {026216} (\bibinfo {year} {2003})}\BibitemShut {NoStop}%
\bibitem [{\citenamefont {Watanabe}\ \emph {et~al.}(1996)\citenamefont
  {Watanabe}, \citenamefont {van~der Zant}, \citenamefont {Strogatz},\ and\
  \citenamefont {Orlando}}]{watanabe1996dynamics}%
  \BibitemOpen
  \bibfield  {author} {\bibinfo {author} {\bibfnamefont {Shinya}\ \bibnamefont
  {Watanabe}}, \bibinfo {author} {\bibfnamefont {Herre~SJ}\ \bibnamefont
  {van~der Zant}}, \bibinfo {author} {\bibfnamefont {Steven~H}\ \bibnamefont
  {Strogatz}}, \ and\ \bibinfo {author} {\bibfnamefont {Terry~P}\ \bibnamefont
  {Orlando}},\ }\bibfield  {title} {\enquote {\bibinfo {title} {Dynamics of
  circular arrays of josephson junctions and the discrete sine-gordon
  equation},}\ }\href@noop {} {\bibfield  {journal} {\bibinfo  {journal}
  {Physica D: Nonlinear Phenomena}\ }\textbf {\bibinfo {volume} {97}},\
  \bibinfo {pages} {429--470} (\bibinfo {year} {1996})}\BibitemShut {NoStop}%
\bibitem [{\citenamefont {Wiley}\ \emph {et~al.}(2006)\citenamefont {Wiley},
  \citenamefont {Strogatz},\ and\ \citenamefont {Girvan}}]{wiley2006size}%
  \BibitemOpen
  \bibfield  {author} {\bibinfo {author} {\bibfnamefont {Daniel~A}\
  \bibnamefont {Wiley}}, \bibinfo {author} {\bibfnamefont {Steven~H}\
  \bibnamefont {Strogatz}}, \ and\ \bibinfo {author} {\bibfnamefont {Michelle}\
  \bibnamefont {Girvan}},\ }\bibfield  {title} {\enquote {\bibinfo {title} {The
  size of the sync basin},}\ }\href@noop {} {\bibfield  {journal} {\bibinfo
  {journal} {Chaos: An Interdisciplinary Journal of Nonlinear Science}\
  }\textbf {\bibinfo {volume} {16}} (\bibinfo {year} {2006})}\BibitemShut
  {NoStop}%
\bibitem [{\citenamefont {Ochab}\ and\ \citenamefont {Gora}(2009)}]{article}%
  \BibitemOpen
  \bibfield  {author} {\bibinfo {author} {\bibfnamefont {Jeremi}\ \bibnamefont
  {Ochab}}\ and\ \bibinfo {author} {\bibfnamefont {Pawel}\ \bibnamefont
  {Gora}},\ }\bibfield  {title} {\enquote {\bibinfo {title} {Synchronization of
  coupled oscillators in a local one-dimensional kuramoto model},}\ }\href@noop
  {} {\bibfield  {journal} {\bibinfo  {journal} {Acta Physica Polonica B,
  Proceedings Supplement}\ }\textbf {\bibinfo {volume} {3}} (\bibinfo {year}
  {2009})}\BibitemShut {NoStop}%
\bibitem [{\citenamefont {Tilles}\ \emph {et~al.}(2011)\citenamefont {Tilles},
  \citenamefont {Ferreira},\ and\ \citenamefont
  {Cerdeira}}]{tilles2011multistable}%
  \BibitemOpen
  \bibfield  {author} {\bibinfo {author} {\bibfnamefont {Paulo~FC}\
  \bibnamefont {Tilles}}, \bibinfo {author} {\bibfnamefont {Fernando~F}\
  \bibnamefont {Ferreira}}, \ and\ \bibinfo {author} {\bibfnamefont {Hilda~A}\
  \bibnamefont {Cerdeira}},\ }\bibfield  {title} {\enquote {\bibinfo {title}
  {Multistable behavior above synchronization in a locally coupled kuramoto
  model},}\ }\href@noop {} {\bibfield  {journal} {\bibinfo  {journal} {Physical
  Review E—Statistical, Nonlinear, and Soft Matter Physics}\ }\textbf
  {\bibinfo {volume} {83}},\ \bibinfo {pages} {066206} (\bibinfo {year}
  {2011})}\BibitemShut {NoStop}%
\bibitem [{\citenamefont {D{\'e}nes}\ \emph {et~al.}(2019)\citenamefont
  {D{\'e}nes}, \citenamefont {S{\'a}ndor},\ and\ \citenamefont
  {N{\'e}da}}]{denes2019pattern}%
  \BibitemOpen
  \bibfield  {author} {\bibinfo {author} {\bibfnamefont {K{\'a}roly}\
  \bibnamefont {D{\'e}nes}}, \bibinfo {author} {\bibfnamefont {Bulcs{\'u}}\
  \bibnamefont {S{\'a}ndor}}, \ and\ \bibinfo {author} {\bibfnamefont
  {Zolt{\'a}n}\ \bibnamefont {N{\'e}da}},\ }\bibfield  {title} {\enquote
  {\bibinfo {title} {Pattern selection in a ring of kuramoto oscillators},}\
  }\href@noop {} {\bibfield  {journal} {\bibinfo  {journal} {Communications in
  Nonlinear Science and Numerical Simulations}\ }\textbf {\bibinfo {volume}
  {78}},\ \bibinfo {pages} {104868} (\bibinfo {year} {2019})}\BibitemShut
  {NoStop}%
\bibitem [{\citenamefont {Denes}\ \emph {et~al.}(2019)\citenamefont {Denes},
  \citenamefont {Sandor},\ and\ \citenamefont
  {Neda}}]{denes2019predictability}%
  \BibitemOpen
  \bibfield  {author} {\bibinfo {author} {\bibfnamefont {Karoly}\ \bibnamefont
  {Denes}}, \bibinfo {author} {\bibfnamefont {Bulcsu}\ \bibnamefont {Sandor}},
  \ and\ \bibinfo {author} {\bibfnamefont {Zoltan}\ \bibnamefont {Neda}},\
  }\bibfield  {title} {\enquote {\bibinfo {title} {On the predictability of the
  final state in a ring of kuramoto rotators},}\ }\href@noop {} {\bibfield
  {journal} {\bibinfo  {journal} {Rom. Rep. Phys}\ }\textbf {\bibinfo {volume}
  {71}},\ \bibinfo {pages} {108} (\bibinfo {year} {2019})}\BibitemShut
  {NoStop}%
\bibitem [{\citenamefont {Kerszberg}\ and\ \citenamefont
  {Zippelius}(1990)}]{kerszberg1990synchronization}%
  \BibitemOpen
  \bibfield  {author} {\bibinfo {author} {\bibfnamefont {Michel}\ \bibnamefont
  {Kerszberg}}\ and\ \bibinfo {author} {\bibfnamefont {Annette}\ \bibnamefont
  {Zippelius}},\ }\bibfield  {title} {\enquote {\bibinfo {title}
  {Synchronization in neural assemblies},}\ }\href@noop {} {\bibfield
  {journal} {\bibinfo  {journal} {Physica Scripta}\ }\textbf {\bibinfo {volume}
  {1990}},\ \bibinfo {pages} {54} (\bibinfo {year} {1990})}\BibitemShut
  {NoStop}%
\bibitem [{\citenamefont {Waibel}\ \emph {et~al.}(2013)\citenamefont {Waibel},
  \citenamefont {Hanazawa}, \citenamefont {Hinton}, \citenamefont {Shikano},\
  and\ \citenamefont {Lang}}]{waibel2013phoneme}%
  \BibitemOpen
  \bibfield  {author} {\bibinfo {author} {\bibfnamefont {Alexander}\
  \bibnamefont {Waibel}}, \bibinfo {author} {\bibfnamefont {Toshiyuki}\
  \bibnamefont {Hanazawa}}, \bibinfo {author} {\bibfnamefont {Geoffrey}\
  \bibnamefont {Hinton}}, \bibinfo {author} {\bibfnamefont {Kiyohiro}\
  \bibnamefont {Shikano}}, \ and\ \bibinfo {author} {\bibfnamefont {Kevin~J}\
  \bibnamefont {Lang}},\ }\bibfield  {title} {\enquote {\bibinfo {title}
  {Phoneme recognition using time-delay neural networks},}\ }in\ \href@noop {}
  {\emph {\bibinfo {booktitle} {Backpropagation}}}\ (\bibinfo  {publisher}
  {Psychology Press},\ \bibinfo {year} {2013})\ pp.\ \bibinfo {pages}
  {35--61}\BibitemShut {NoStop}%
\bibitem [{\citenamefont {Kozyreff}\ \emph {et~al.}(2000)\citenamefont
  {Kozyreff}, \citenamefont {Vladimirov},\ and\ \citenamefont
  {Mandel}}]{kozyreff2000global}%
  \BibitemOpen
  \bibfield  {author} {\bibinfo {author} {\bibfnamefont {Gregory}\ \bibnamefont
  {Kozyreff}}, \bibinfo {author} {\bibfnamefont {AG}~\bibnamefont
  {Vladimirov}}, \ and\ \bibinfo {author} {\bibfnamefont {Paul}\ \bibnamefont
  {Mandel}},\ }\bibfield  {title} {\enquote {\bibinfo {title} {Global coupling
  with time delay in an array of semiconductor lasers},}\ }\href@noop {}
  {\bibfield  {journal} {\bibinfo  {journal} {Physical Review Letters}\
  }\textbf {\bibinfo {volume} {85}},\ \bibinfo {pages} {3809} (\bibinfo {year}
  {2000})}\BibitemShut {NoStop}%
\bibitem [{\citenamefont {Reddy}\ \emph {et~al.}(2000)\citenamefont {Reddy},
  \citenamefont {Sen},\ and\ \citenamefont {Johnston}}]{reddy2000experimental}%
  \BibitemOpen
  \bibfield  {author} {\bibinfo {author} {\bibfnamefont {DV~Ramana}\
  \bibnamefont {Reddy}}, \bibinfo {author} {\bibfnamefont {Abhijit}\
  \bibnamefont {Sen}}, \ and\ \bibinfo {author} {\bibfnamefont {George~L}\
  \bibnamefont {Johnston}},\ }\bibfield  {title} {\enquote {\bibinfo {title}
  {Experimental evidence of time-delay-induced death in coupled limit-cycle
  oscillators},}\ }\href@noop {} {\bibfield  {journal} {\bibinfo  {journal}
  {Physical Review Letters}\ }\textbf {\bibinfo {volume} {85}},\ \bibinfo
  {pages} {3381} (\bibinfo {year} {2000})}\BibitemShut {NoStop}%
\bibitem [{\citenamefont {Yeung}\ and\ \citenamefont
  {Strogatz}(1999)}]{yeung1999time}%
  \BibitemOpen
  \bibfield  {author} {\bibinfo {author} {\bibfnamefont {MK~Stephen}\
  \bibnamefont {Yeung}}\ and\ \bibinfo {author} {\bibfnamefont {Steven~H}\
  \bibnamefont {Strogatz}},\ }\bibfield  {title} {\enquote {\bibinfo {title}
  {Time delay in the kuramoto model of coupled oscillators},}\ }\href@noop {}
  {\bibfield  {journal} {\bibinfo  {journal} {Physical review letters}\
  }\textbf {\bibinfo {volume} {82}},\ \bibinfo {pages} {648} (\bibinfo {year}
  {1999})}\BibitemShut {NoStop}%
\bibitem [{\citenamefont {Taher}\ \emph {et~al.}(2019)\citenamefont {Taher},
  \citenamefont {Olmi},\ and\ \citenamefont {Sch{\"o}ll}}]{taher2019enhancing}%
  \BibitemOpen
  \bibfield  {author} {\bibinfo {author} {\bibfnamefont {Halgurd}\ \bibnamefont
  {Taher}}, \bibinfo {author} {\bibfnamefont {Simona}\ \bibnamefont {Olmi}}, \
  and\ \bibinfo {author} {\bibfnamefont {Eckehard}\ \bibnamefont
  {Sch{\"o}ll}},\ }\bibfield  {title} {\enquote {\bibinfo {title} {Enhancing
  power grid synchronization and stability through time-delayed feedback
  control},}\ }\href@noop {} {\bibfield  {journal} {\bibinfo  {journal}
  {Physical Review E}\ }\textbf {\bibinfo {volume} {100}},\ \bibinfo {pages}
  {062306} (\bibinfo {year} {2019})}\BibitemShut {NoStop}%
\bibitem [{\citenamefont {Wernecke}\ \emph {et~al.}(2019)\citenamefont
  {Wernecke}, \citenamefont {S{\'a}ndor},\ and\ \citenamefont
  {Gros}}]{wernecke2019chaos}%
  \BibitemOpen
  \bibfield  {author} {\bibinfo {author} {\bibfnamefont {Hendrik}\ \bibnamefont
  {Wernecke}}, \bibinfo {author} {\bibfnamefont {Bulcs{\'u}}\ \bibnamefont
  {S{\'a}ndor}}, \ and\ \bibinfo {author} {\bibfnamefont {Claudius}\
  \bibnamefont {Gros}},\ }\bibfield  {title} {\enquote {\bibinfo {title} {Chaos
  in time delay systems, an educational review},}\ }\href@noop {} {\bibfield
  {journal} {\bibinfo  {journal} {Physics Reports}\ }\textbf {\bibinfo {volume}
  {824}},\ \bibinfo {pages} {1--40} (\bibinfo {year} {2019})}\BibitemShut
  {NoStop}%
\bibitem [{\citenamefont {Choi}\ \emph {et~al.}(2000)\citenamefont {Choi},
  \citenamefont {Kim}, \citenamefont {Kim},\ and\ \citenamefont
  {Hong}}]{choi2000synchronization}%
  \BibitemOpen
  \bibfield  {author} {\bibinfo {author} {\bibfnamefont {MY}~\bibnamefont
  {Choi}}, \bibinfo {author} {\bibfnamefont {HJ}~\bibnamefont {Kim}}, \bibinfo
  {author} {\bibfnamefont {D}~\bibnamefont {Kim}}, \ and\ \bibinfo {author}
  {\bibfnamefont {H}~\bibnamefont {Hong}},\ }\bibfield  {title} {\enquote
  {\bibinfo {title} {Synchronization in a system of globally coupled
  oscillators with time delay},}\ }\href@noop {} {\bibfield  {journal}
  {\bibinfo  {journal} {Physical Review E}\ }\textbf {\bibinfo {volume} {61}},\
  \bibinfo {pages} {371} (\bibinfo {year} {2000})}\BibitemShut {NoStop}%
\bibitem [{\citenamefont {Ameli}\ \emph {et~al.}(2021)\citenamefont {Ameli},
  \citenamefont {Karimian},\ and\ \citenamefont {Shahbazi}}]{ameli2021time}%
  \BibitemOpen
  \bibfield  {author} {\bibinfo {author} {\bibfnamefont {Sara}\ \bibnamefont
  {Ameli}}, \bibinfo {author} {\bibfnamefont {Maryam}\ \bibnamefont
  {Karimian}}, \ and\ \bibinfo {author} {\bibfnamefont {Farhad}\ \bibnamefont
  {Shahbazi}},\ }\bibfield  {title} {\enquote {\bibinfo {title} {Time-delayed
  kuramoto model in the watts--strogatz small-world networks},}\ }\href@noop {}
  {\bibfield  {journal} {\bibinfo  {journal} {Chaos: An Interdisciplinary
  Journal of Nonlinear Science}\ }\textbf {\bibinfo {volume} {31}} (\bibinfo
  {year} {2021})}\BibitemShut {NoStop}%
\bibitem [{\citenamefont {Earl}\ and\ \citenamefont
  {Strogatz}(2003)}]{earl2003synchronization}%
  \BibitemOpen
  \bibfield  {author} {\bibinfo {author} {\bibfnamefont {Matthew~G}\
  \bibnamefont {Earl}}\ and\ \bibinfo {author} {\bibfnamefont {Steven~H}\
  \bibnamefont {Strogatz}},\ }\bibfield  {title} {\enquote {\bibinfo {title}
  {Synchronization in oscillator networks with delayed coupling: A stability
  criterion},}\ }\href@noop {} {\bibfield  {journal} {\bibinfo  {journal}
  {Physical Review E}\ }\textbf {\bibinfo {volume} {67}},\ \bibinfo {pages}
  {036204} (\bibinfo {year} {2003})}\BibitemShut {NoStop}%
\bibitem [{\citenamefont {D{\'e}nes}\ \emph {et~al.}(2021)\citenamefont
  {D{\'e}nes}, \citenamefont {S{\'a}ndor},\ and\ \citenamefont
  {N{\'e}da}}]{denes2021synchronization}%
  \BibitemOpen
  \bibfield  {author} {\bibinfo {author} {\bibfnamefont {K{\'a}roly}\
  \bibnamefont {D{\'e}nes}}, \bibinfo {author} {\bibfnamefont {Bulcs{\'u}}\
  \bibnamefont {S{\'a}ndor}}, \ and\ \bibinfo {author} {\bibfnamefont
  {Zolt{\'a}n}\ \bibnamefont {N{\'e}da}},\ }\bibfield  {title} {\enquote
  {\bibinfo {title} {Synchronization patterns in rings of time-delayed kuramoto
  oscillators},}\ }\href@noop {} {\bibfield  {journal} {\bibinfo  {journal}
  {Communications in Nonlinear Science and Numerical Simulation}\ }\textbf
  {\bibinfo {volume} {93}},\ \bibinfo {pages} {105505} (\bibinfo {year}
  {2021})}\BibitemShut {NoStop}%
\bibitem [{\citenamefont {Matias}\ \emph {et~al.}(1997)\citenamefont {Matias},
  \citenamefont {P{\'e}rez-Mu{\~n}uzuri}, \citenamefont {Lorenzo},
  \citenamefont {Marino},\ and\ \citenamefont
  {P{\'e}rez-Villar}}]{matias1997observation}%
  \BibitemOpen
  \bibfield  {author} {\bibinfo {author} {\bibfnamefont {MA}~\bibnamefont
  {Matias}}, \bibinfo {author} {\bibfnamefont {V}~\bibnamefont
  {P{\'e}rez-Mu{\~n}uzuri}}, \bibinfo {author} {\bibfnamefont {MN}~\bibnamefont
  {Lorenzo}}, \bibinfo {author} {\bibfnamefont {IP}~\bibnamefont {Marino}}, \
  and\ \bibinfo {author} {\bibfnamefont {V}~\bibnamefont {P{\'e}rez-Villar}},\
  }\bibfield  {title} {\enquote {\bibinfo {title} {Observation of a fast
  rotating wave in rings of coupled chaotic oscillators},}\ }\href@noop {}
  {\bibfield  {journal} {\bibinfo  {journal} {Physical review letters}\
  }\textbf {\bibinfo {volume} {78}},\ \bibinfo {pages} {219} (\bibinfo {year}
  {1997})}\BibitemShut {NoStop}%
\bibitem [{\citenamefont {Esfahani}\ \emph {et~al.}(2012)\citenamefont
  {Esfahani}, \citenamefont {Shahbazi},\ and\ \citenamefont
  {Samani}}]{esfahani2012noise}%
  \BibitemOpen
  \bibfield  {author} {\bibinfo {author} {\bibfnamefont {Reihaneh~Kouhi}\
  \bibnamefont {Esfahani}}, \bibinfo {author} {\bibfnamefont {Farhad}\
  \bibnamefont {Shahbazi}}, \ and\ \bibinfo {author} {\bibfnamefont
  {Keivan~Aghababaei}\ \bibnamefont {Samani}},\ }\bibfield  {title} {\enquote
  {\bibinfo {title} {Noise-induced synchronization in small world networks of
  phase oscillators},}\ }\href@noop {} {\bibfield  {journal} {\bibinfo
  {journal} {Physical Review E}\ }\textbf {\bibinfo {volume} {86}},\ \bibinfo
  {pages} {036204} (\bibinfo {year} {2012})}\BibitemShut {NoStop}%
\bibitem [{\citenamefont {Kemeth}\ \emph {et~al.}(2016)\citenamefont {Kemeth},
  \citenamefont {Haugland}, \citenamefont {Schmidt}, \citenamefont
  {Kevrekidis},\ and\ \citenamefont {Krischer}}]{kemeth2016classification}%
  \BibitemOpen
  \bibfield  {author} {\bibinfo {author} {\bibfnamefont {Felix~P}\ \bibnamefont
  {Kemeth}}, \bibinfo {author} {\bibfnamefont {Sindre~W}\ \bibnamefont
  {Haugland}}, \bibinfo {author} {\bibfnamefont {Lennart}\ \bibnamefont
  {Schmidt}}, \bibinfo {author} {\bibfnamefont {Ioannis~G}\ \bibnamefont
  {Kevrekidis}}, \ and\ \bibinfo {author} {\bibfnamefont {Katharina}\
  \bibnamefont {Krischer}},\ }\bibfield  {title} {\enquote {\bibinfo {title} {A
  classification scheme for chimera states},}\ }\href@noop {} {\bibfield
  {journal} {\bibinfo  {journal} {Chaos: An Interdisciplinary Journal of
  Nonlinear Science}\ }\textbf {\bibinfo {volume} {26}} (\bibinfo {year}
  {2016})}\BibitemShut {NoStop}%
\end{thebibliography}%

\end{document}